\documentclass[onecolumn,fleqn,usenatbib]{mnras}

\usepackage{newtxtext,newtxmath}

\usepackage[T1]{fontenc}

\DeclareRobustCommand{\VAN}[3]{#2}
\let\VANthebibliography\thebibliography
\def\thebibliography{\DeclareRobustCommand{\VAN}[3]{##3}\VANthebibliography}

\usepackage{graphicx}	
\usepackage{amsmath}	
\usepackage{soul}
\usepackage{xcolor}
\usepackage{orcidlink} 
\usepackage{float}
\usepackage{caption}
\usepackage{subcaption}

\def\rg{r_{\rm g}}
\def\rs{r_{\rm s}}
\def\mbh{M_{\rm {bh}}}
\def\ve{\rm v}
\def\vr{v_{\rm r}}
\def\vinj{v_{\rm inj}}

\def\hinj{H_{\rm inj}}
\def\vz{v_{\rm z}}

\def\vp{v_{\rm \phi}}

\def\tvir{T_{\rm vir}}
\def\thvir{\Theta_{\rm vir}}

\def\tob{T_{\rm ob}}
\def\thob{\Theta_{\rm ob}}

\def\rhinj{\rho_{\rm inj}}

\def\lmk{\lambda_{\rm K}}
\def\lmdob{\lambda_{\rm ob}}
\def\lmdinj{\lambda_{\rm inj}}

\def\lsim{\lower.5ex\hbox{$\; \buildrel < \over \sim \;$}}
\def\gsim{\lower.5ex\hbox{$\; \buildrel > \over \sim \;$}}
\def\simeq{\lower.3ex\hbox{$\; \buildrel \sim \over - \;$}}
\def\rinj{r_{\rm inj}}
\def\rob{r_{\rm ob}}

\def\cs{c_{\rm s}}
\def\csinj{c_{\rm sinj}}

\def\sun{M_\odot}

\def\medd{\dot{M}_{\rm Edd}}

\def\tt{\tilde{t}}

\def\be{B_{\rm ob}}

\def\hm{{hot-mode} }
\def\cm{{cold-mode} }

\title[Simulation of hot accretion flows]{Simulation of advective accretion flows around black holes under various outer boundary conditions}
\author[Kumar \& Lee]
{Rajiv Kumar$^{1}$\thanks{E-mail: rajivbhu09@gmail.com}, 
Seong-Jae Lee$^{2}$\thanks{E-mail: seong@chungbuk.ac.kr} 
\\
$^{1}$Harish-Chandra Research Institute, HBNI, Jhunsi, Prayagraj 211019, India\\
$^{2}$School of Science Education, Chungbuk National University, Chungbuk 28644, S. Korea
}

\date{Accepted XXX. Received YYY; in original form ZZZ}

\begin{document}
\label{firstpage}
\pagerange{\pageref{firstpage}--\pageref{lastpage}}
\maketitle

\begin{abstract}
We simulated various accretion disc structures with viscous hydrodynamic (HD) flow around a black hole (BH). We found that the structure of the accretion disc is significantly influenced by changes in the physical parameters of the initial inflowing gases. These physical parameters can be called outer boundary conditions (OBCs) at the outer-accretion boundary and represented on an OBC plane, which is primarily divided into hot-mode and cold-mode inflowing gases. We found smooth and shocked flows in the simulation, which follow the semi-analytical solutions with their OBCs. Interestingly, we observed that certain types of OBCs can produce shocks with jet-like features in the accretion flow. However, other OBCs can produce smooth or shock-free accretion flow, which may or may not have outflows. The smooth flows can display both the lowest and highest angular momentum distributions among the advective flows, depending on the OBCs. Additionally, the nature of the accretion flows can be either steady or quasi-steady, also influenced by the OBCs. We also observed that solutions corresponding to hot-mode gases have a greater tendency to generate outflows compared to those with the cold-mode. Therefore, this qualitative study of OBCs is crucial for understanding accretion physics, which can aid in modeling accretion discs, similar to the quantitative studies (which involve only changing mass accretion rates) of the inflowing gases. Thus, we assert that an accretion model should be based on both the qualitative and quantitative aspects of the initial inflowing gases.
\end{abstract}




\section{Introduction}\label{sec:intro}
An accretion flow powers many astrophysical objects, including black hole X-ray binaries (BH XRBs), tidal disruption events (TDEs), active galactic nuclei (AGNs), and others. 
The BH XRBs usually possess two types of spectral states: a soft state and a hard state \citep{fbg04,rm06,bmm11,nb18}. The AGN's spectra also contain soft and hard photons. The soft state features a multicolour blackbody thermal spectrum, while the hard state has a non-thermal power-law high-energy component. 
Additionally, these accreting sources exhibit various time scales of variability, which can be either quasi-periodic or aperiodic. The spectral states in BH XRBs can be illustrated as a q-diagram in the hardness-intensity diagram (HID) with connecting intermediate spectral states \citep{fbg04}. The origins of the two spectral states differ significantly. The soft state can arise from the non-advective cool Keplerian disc \citep{ss73}, also known as the Shakura-Sunyaev disc.  
Conversely, the hard state can be produced by very hot gas, which may consist of an advective hot sub-Keplerian flow or a corona-like structure near the BH. Advective hot flows come in wide varieties, as noted in the literature, including shocked flows \citep{c89}, Bondi/Bondi-type flows \citep{B52,m72}, advection-dominated accretion flows (ADAFs) \citep{ny94,nkh97}, and ADAF-thick flows \citep{lgy99}. Therefore, two models based on the configurations of the two hybrid flows (hot advective and cool non-advective) are employed to explain the different spectral states of the BH XRBs and AGNs. Those models are known as the sandwich geometry model \citep{ct95} and the two-zone radial geometry model \citep{emn97}. Both models considered only one type of advective solution, whether shocked flow or smooth ADAF flow. Interestingly, all these advective flows stem from the same set of fluid equations of motion, albeit with different OBCs in the semi-analytical studies \citep{y99,ky21,r25}. Thus, we refer to these solutions collectively as general advective hot accretion flows (hereafter referred to as advective flows). \citet{ky21} has shown that the global ADAFs can only be generated when the local specific energy (denoted as $\be$)$<0$ of inflowing gases at the outer accretion boundary, and other advective flows can be generated when $\be>0$.
The corresponding OBCs can be represented in the OBC-plane \citep{kg19,ky21,r25}, with further details about the OBCs written in section \ref{sec:obcs} of this study. 
One of the primary objectives of the present simulation study is to explore various advective flows using different OBCs as found in the semi-analytical studies.

Since the hard state of the BH XRBs is highly variable compared to the soft state, the outflows/jets are associated with it. 
It is found that a hard state can be generated from the hot advective flows, so whether these flows exhibit an outflowing nature is another objective of this simulation study that involves changing the OBCs. For instances, some TDEs exhibit jet emissions; this phenomenon is not observed in most cases \citep[reference therein]{dlc21}. 
{ Interestingly, the state transitions in TDEs are also predicted to be similar to those in BH XRBs \citep{bkc26}. These TDEs can exhibit both thermal and non-thermal radiations, suggesting that a TDE could have two flows: a cool super-Eddington flow dominating in the early phase and a hot sub-Eddington flow dominating in the late phase. However, the disk structures of TDEs are not very clear and debatable. The TDEs in the late phase can exhibit outflows \citep{am26}. Therefore, this study can also have the potential to shed light on the generation of outflows in TDEs.}  
Furthermore, observational studies have shown that the shape of the HID diagram in BH XRBs can deviate from one outburst to another 
{ in a same BH source, like, recurrent outbursts of GX 339-4 in different epoch has shown different shape and sizes of HIDs in different outbursts \citep{dmb08,drd21,ay22}}.
Moreover, the luminosities and jet strength (strong or weak jets) of a same source can fluctuate from one outburst to another. 
{ Since the inner BH conditions of a particular BH X-ray binary remain unchanged over a sufficiently long period, the observed variations in luminosities and jet strength across different outburst cycles can be attributed to the OBCs.} Therefore, { the study of OBCs is very important, and} we believe these observed variations can be attributed to the qualitative and quantitative nature of the initial inflowing gases at the outer accretion boundary. 
This study focuses on the qualitative characteristics of the inflowing gases at the outer accretion boundary, such as initial angular momentum, initial temperature, initial velocity of the inflowing gases, and location of the outer boundary, which we refer to as the parameters of the OBCs. 
Furthermore, numerous studies have indicated that {advective} flows possess sub-Keplerian angular momentum distributions and move significantly faster than the Keplerian disc flow. Consequently, {advective} flows may also produce rapid variabilities in the hard state of the sources.   
As we know, most of the observed properties (some mentioned above) of the sources have come from the inner part of the disc (say maybe $<200\rs$). Since, the nature of the inner part of the disc around the BH can be dependent on the OBCs, as seen in the semi-analytical studies. Therefore, it is crucial to investigate the hot advective flows in relation to the OBCs. For the time being, this simulation study is based on the qualitative nature of the OBC, and more details about the OBCs are presented in the section \ref{sec:obcs}.

Many simulation studies have explored accretion flows around the BHs in the HD regime 
\citep[references therein]{msc96,lmc98,lrc11,ybw12,ywb12,dcn14,lck16,smg22,dcj24} as well as in magnetohydrodynamics (MHD) regimes 
\citep[references therein]{gbc20,zyk20,dvf21,ybw12,ywb12,mfa22,apo24,Zhao_2024}. Most simulation studies have utilised an initial background setup that concentrates on one type of solution or another, focusing on specific features of the accreting sources. For example, many studies employed a torus-like structure setup using various initial profiles of physical quantities (e.g., gas density, pressure, or temperature with initial angular momentum) around an equatorial plane in the simulation box 
\citep{mdm25,apo24,mfa22,snt15,nsp12,ybw12,ywb12}, while some studies used an initial background setup across the entire simulation box 
\citep{zyk20,wby22,zby22,Zhao_2024}. In the present study, we applied initial input parameters or simulation boundary conditions (BCs) at the outer boundary of the simulation box, and details are discussed in subsection \ref{subsec:SBC}. The initial input parameters are taken from the semi-analytical solutions (Equations are discussed in Appendix \ref{sec:tsss}). This type of input conditions in the simulations provides some important opportunities, such as testing the simulation results with semi-analytical studies, helping us to investigate different disc structures, and allowing us to connect the simulation BCs with the theoretical/physical OBCs of the accretion system. However, the same method for determining input conditions was also utilised in previous HD and MHD simulation studies with non-relativistic \citep{msc96,dcn14,lck16,gbc20,soa21} and relativistic approaches \citep{g25,dum25}, although boundary condition selection was limited and did not fully explore all kinds of OBCs and corresponding disc structures. Moreover, they used inviscid semi-analytical solutions for input parameters in their simulations. In the present study, we used viscous semi-analytical solutions and systematically explored the OBCs of the inflowing gases. None of the previous studies have categorised the OBCs or examined them in relation to the qualitative nature of the inflowing gases at the outer accretion boundary. In this regard, the current simulation study represents a new initiative to discuss the OBCs of the accreting system's external environment and their corresponding effects on the accretion flows around the BHs.

In this study, we analyse the accretion disc structures that emerge from simulations with varying OBCs (such as initial angular momentum and initial temperature). Moreover, we will also endeavour to relate our simulation results to some observed properties of the sources. The structure of this paper is as follows: Section \ref{sec:eqns} provides a brief introduction to the simulation code and assumptions; Section \ref{sec:obcs} introduces the OBCs and simulation BCs; Section \ref{sec:result} presents the results; and in the final Section \ref{sec:sumup}, we offer a summary and discussions.

\section{Brief Introduction of Simulation code} \label{sec:eqns}
This numerical simulation study has been done on the 2.5-dimensional cylindrical geometry based on the basic HD equations and the code of \cite{lck16}.
The code used the Lagrangian Total Variation Diminishing (LTVD) code with a remap routine method to handle efficiently the angular momentum distribution. The LTVD code consists of Lagrangian and TVD schemes designed by \cite{ryc95} based on \cite{h83}. {It makes the angular momentum follow as accurately as possible and can detect shocks sharply, so it is very useful to observe the angular momentum distribution, shock capture, shock collision, etc.} 
In particular, it is also very helpful in analysing the structure of the accretion disc around a BH depending on the properties of the inflow.
This code has been tested with the accretion flow result of other simulation work \citep{msc96} by regenerating the accretion flow with the same boundary conditions. 
The present study has focused on the investigation of viscous accretion flows based on different physical properties of the inflowing gas at the outer boundary and checked the predictions of the semi-analytical studies and corresponding OBCs. We have also compared the simulation results with low and high resolutions, which are generated from the same types of OBCs.
 Usually, the higher resolution results are good for the shock capturing and comparing the shock location with the semi-analytical accretion solutions. 
 These tests can provide the robustness of the solutions and the suitability of the simulation code. 
The details of the numerical calculation and equations of the LTVD code are described in the previous studies \cite{lck16,lrc11}. For the convenience of the reader, we have briefly discussed the basic equations and code schemes here as follows.\\
Accreting fluid has a tendency to settle into a disc shape due to its angular momentum and forms an axis-symmetric system. Therefore, we assumed a rotating and axis-symmetric flow in cylindrical geometry $(r,\phi,z)$.
Following the \cite{lck16}, the basic equations used for this simulation study can be described by, the mass conservation equation is,
\begin{equation}
\centering
\frac{\partial\rho}{\partial t}+\frac{1}{r}\frac{\partial(r\rho\vr)}{\partial r}+\frac{\partial(\rho\vz)}{\partial z}=0,
\label{mc.eq}
\end{equation}
 the components of the momentum conservation equations are, 
 \begin{eqnarray}
\frac{\partial(\rho\vr)}{\partial t}+\frac{1}{r}\frac{\partial(r\rho\vr^2)}{\partial r}+\frac{\partial(\rho\vr\vz)}{\partial z}+\frac{\partial p}{\partial r}=-\rho\frac{\partial\Phi}{\partial r}+\rho\frac{\lambda^2}{r^3},\label{mcr.eq}\\
\frac{\partial(\rho\vp)}{\partial t}+\frac{1}{r}\frac{(r\rho\vp\vr)}{\partial r}+\frac{\partial(\rho\vp\vz)}{\partial z}=\frac{1}{r^2}\frac{\partial}{\partial r}(r^2\sigma_{r\phi})+r\frac{\partial}{\partial z}\left(\frac{\sigma_{z\phi}}{r}\right), \label{mcp.eq}\\
\frac{\partial(\rho\vz)}{\partial t}+\frac{1}{r}\frac{\partial(r\rho\vr\vz)}{\partial r}+\frac{\partial(\rho\vz^2)}{\partial z}+\frac{\partial p}{\partial z}=-\rho\frac{\partial\Phi}{\partial z},
\label{mcz.eq}
\end{eqnarray}
and the energy conservation equation is,
\begin{equation}
\frac{\partial\epsilon}{\partial t}+\frac{1}{r}\frac{\partial(r\epsilon\vr)}{\partial r}+\frac{\partial(\epsilon\vz)}{\partial z}+\frac{1}{r}\frac{\partial(rp\vr)}{\partial r}+\frac{\partial(p\vz)}{\partial z}=\frac{1}{r}\frac{\partial(r\vp\sigma_{r\phi})}{\partial r}+\frac{\partial(\vp\sigma_{z\phi})}{\partial z}-\rho\vr\frac{\partial\Phi}{\partial r}-\rho\vz\frac{\partial\Phi}{\partial z},
\label{ec.eq}
\end{equation}
where, $\rho, p, \vr, \vp, \vz$, and $\lambda$ are the gas density, gas pressure, radial velocity, azimuthal velocity, vertical velocity, and specific angular momentum, respectively. Here, $\epsilon=\rho(\vr^2+\vp^2+\vz^2)/2+\rho e$ is total energy density, and $e=p/[\rho(\gamma-1)]$ is specific internal energy of gas, where, $\gamma$ is the adiabatic index. With following \citet{spb99,kg18}, we have assumed that only two components of the viscous stress tensor are effective in the $r-z$ plane of the disc, which are $\sigma_{r\phi}=r\mu(\partial\Omega/\partial r)$ and $\sigma_{z\phi}=\mu(\partial\vp/\partial z)$, and $\Omega=\vp/r=\lambda/r^2$, where, $\mu=\alpha\rho(c_s^2/\Omega_K)$ is the dynamical viscosity coefficient, $c_s=\sqrt{\gamma p/\rho}$ is sound speed, and $\Omega_K=\sqrt{(\partial\Phi/\partial r)/r}$ is the Keplerian angular velocity. 
The pseudo-Newtonian gravity potential \citep{pw80} is,
\begin{equation}
\Phi=-\frac{G\mbh}{R-\rg}; ~\mbox{where}~R=\sqrt{r^2+z^2}.
\label{pw.eq}
\end{equation}
Here $\rg=2G\mbh/c^2$ is the gravitational radius, which is equal to the Schwarzschild radius for a non-rotating BH. 
{ This pseudo potential mimics the most of essential properties of the Schwarzschild metric.}
We used the geometrical unit system $2G=\mbh=c=1$, where $G, \mbh$, and $c$ are the universal gravitational constant, the mass of the BH, and the speed of light, respectively.
All quantities are expressed in geometric units, such as $\rg, c, \rg c$, and $\rg/c$, which are the units of length, velocity, specific angular momentum, and time, respectively.

The simulation code is composed of two parts: the hydrodynamic and viscosity parts. The hydrodynamic part is based on the Lagrangian Total Variation Diminishing (TVD) plus remap approach, which preserves angular momentum strictly. 
The viscous parts include the viscous angular momentum transfer and the viscous heating. The viscous angular momentum transfer is updated using the implicit method, ensuring it is free from numerical instabilities related to it. The viscous heating is updated using a second-order explicit method, as it is less susceptible to numerical instabilities.

The hydrodynamic part consists of the Lagrangian step and the remap step. First, in the Lagrangian step, the equations for Lagrangian hydrodynamics are solved on the Lagrangian grid. The upwind schemes are applied to build codes that advance the Lagrangian step using Harten’s TVD scheme, which is an explicit, second-order, finite-difference scheme to solve a hyperbolic system of conservation equations \citep{h83,rok93}. 
In the remap step, the quantities evolved in the Lagrangian grid are redistributed to the Eulerian grid to preserve the spatially fixed grid structure. Before the Lagrangian step, the Lagrangian and Eulerian grid zones coincide. But after the step, the Lagrangian grid zone moves to the updated position; $r^{new}=r^{old}+\bar{v}\Delta t$, $\bar{v}$ is the time-averaged velocity. For the remap, we employ the third-order accurate scheme used in the PPM code \citep{cw84}. 
With the Lagrangian and remap steps, equations  (\ref{mc.eq}-\ref{ec.eq}) are updated in the Eulerian grid, except for the centrifugal force, gravity, and viscosity terms. 
The centrifugal force and gravity terms are calculated separately after the Lagrangian and remap steps, such that
\begin{equation}
v_i^{hydro}=v_i^{lag+remap}+\Delta t(\frac{\lambda_i^{remap}}{r_i^3}-\frac{d\Phi}{dr}|_i)
\end{equation}
Then, the viscosity terms are calculated, as discussed in the following paragraph.

Viscosity has two effects on accretion flows. First, it transfers the angular momentum outward, allowing the matter to accrete inward. At the same time, it acts as friction, which results in viscous heating. Here, the viscous heating energy is fully stored as entropy, since we ignore cooling.
The angular momentum transfer in equation (\ref{mcp.eq}) is described by the viscosity parameter given in \cite{ss73}.
The terms for the angular momentum transfer in radial ($r$) and vertical ($z$) directions in equation (\ref{mcp.eq}) are linear in $\lambda$; it can be solved implicitly. Substituting $(\lambda^{new} +\lambda^{remap})/2$ for $\lambda$, equation (\ref{mcp.eq}) without the advection term becomes 
\begin{equation}
a_i\lambda_{i-1}^{new}+b_i\lambda_i^{new}+c_i\lambda_{i+1}^{new}=-a_i\lambda_{i-1}^{remap}-(b_i-2)\lambda_i^{remap}-c_i\lambda_{i+1}^{remap}
\end{equation}
forming a tridiagonal matrix. Here, $a_i, b_i$, and $c_i$ are given with $\rho, \mu$, and $r$ as well as $\Delta r$ and $\Delta t$. while similarly $a_i^{\prime}, b_i^{\prime}$, and $c_i^{\prime}$ are given with $\rho, \mu$, and $z$ as well as  $\Delta z$ and $\Delta t$.
The tridiagonal matrix can be solved for $\lambda^{new}$ with an appropriate boundary condition \citep{pwt92}. 
It is found in numerical experiments that the explicit treatment for the calculation of the viscous heating term does not cause any numerical problems. Thus, the angular momentum transfer is solved implicitly, while the viscous heating term is solved explicitly.
\section{OBC-plane for inflowing gases}\label{sec:obcs}
BHs do not have their own atmospheres and instead accrete matter from the outside, which forms an accretion disc. Therefore, we believe that the different spectral states with their luminosities and the on/off behaviour of outflows/jets in the sources can be influenced by the physical properties of the inflowing gases in the external environment. These physical properties of the inflowing gases can have quantitative (mass accretion rate) and qualitative in nature at the outer boundary location, which are referred to as the OBCs. 
The qualitative nature of the inflowing gases at the outer boundary can include local temperature (thermal energy), local angular momentum (rotational energy), magnetised/non-magnetised gas, gas composition, and so on. Therefore, the OBCs can also be expressed in terms of the total local specific energy of the gas.
Understanding the OBCs of accreting objects is crucial for both theoretical and numerical studies. BH XRBs, AGNs, and TDEs have the same inner boundary conditions in the form of the BH horizon. However, the OBCs can differ based on factors such as the gas feeding mechanism, disc size, and the qualitative and quantitative nature of the inflowing gases. For instance, 
TDEs feature a small disc ($\sim10^2\rg$) fed by stellar tidal disruptive gas, while AGNs typically have larger discs ($\lsim 10^5\rg$) fed through various means, including torus gas, stellar winds, ISM gas, gas clouds in the BLR, and more. BH XRBs also exhibit disc sizes ($\sim10^{5-7}\rg$) fed by gas from a companion star via winds and Roche-lobe overflow. Thus, the diverse sources of inflowing gases and feeding mechanisms may establish different initial temperatures, angular momentum of the inflowing gases, and other physical properties of gases.
Consequently, we aim to develop a systematic theoretical understanding of these OBCs, where the physical properties of inflowing gases can be parameterised by an initial specific energy parameter at the outer boundary, which includes an initial temperature parameter, an initial angular momentum parameter, an outer boundary location, and an initial inflow velocity parameter. For the first time, we investigated an OBC-plane in terms of the local specific energy parameter versus the outer accretion boundary locations in the accreting objects \citep{kg19,ky21}. 
Through our semi-analytical explorations, we identified possibilities that OBCs can alter the inner disc structure. Our primary objective in the current simulation study is to determine whether the accretion disc structure can change with varying OBCs, as observed in the semi-analytical studies in relativistic \citep{ky21,ck16} and semi-relativistic \citep{k24,kc13,y99} regimes.

\begin{figure}
\begin{center}
 \includegraphics[angle=0, width=0.49\textwidth]{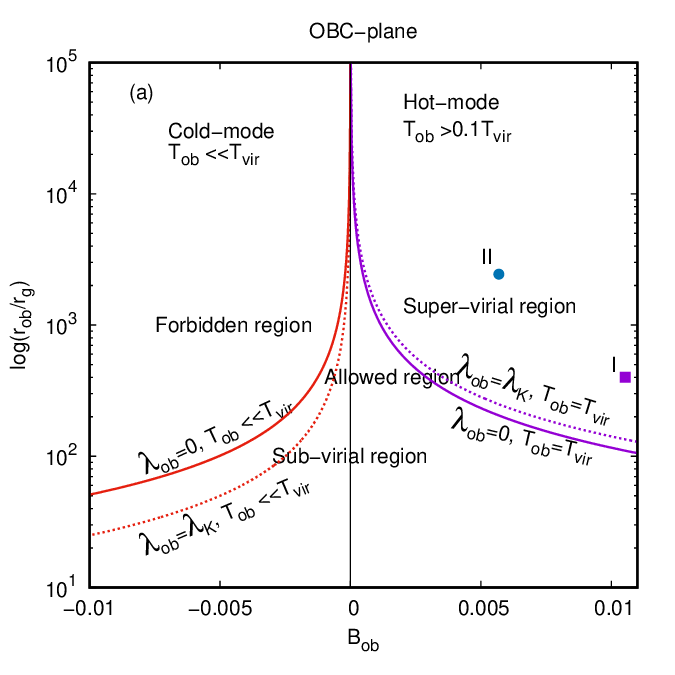}
    \includegraphics[angle=0, width=0.49\textwidth]{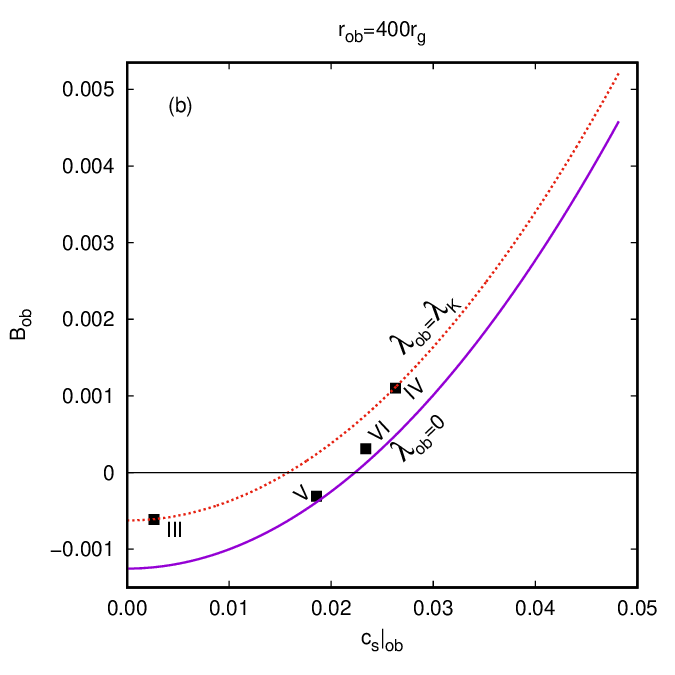}
\caption{A panel (a) represents the OBC-plane of inflowing gases for the accretion around the accreting objects. The Forbidden region represents the energies at the OBLs that are not allowed for the gases around the central objects. The allowed region is divided into cold-mode and hot-mode regions. The hot-mode region comprises sub-virial and super-virial regions. The cold-mode region has only a sub-virial region.
Symbols $\blacksquare$ and $\bullet$ in the super-virial region represent the OBCs, and corresponding solutions, presented in Fig. \ref{Fig3}, and Fig. \ref{Fig8R}, respectively.
A panel (b) represents the energy range of the inflowing gases at $\rob=400\rg$ between the upper dotted line (red colour) and lower solid line (violet colour). 
 Here, $c_s|_{ob}$ is the local sound speed of the inflowing gases at OBL, and it is plotted up to the sound speed corresponding to the virial temperature in both curves.  
 The four $\blacksquare$ symbols denoted in between both curves represent the OBCs of the inflowing gases, and the corresponding solutions are generated and presented in Fig. \ref{Fig4}.}
\label{Fig1}
\end{center}
\end{figure}
It is expected that the BHs or other accreting objects can accrete all kinds of matter available at the accretion boundary, except for local super-Keplerian gas (which exceeds gravitational force), so the accreting gas can have many free parameters, such as initial temperature, initial angular momentum, and more. This simulation study aims to understand the nature of the accretion solutions alongside the corresponding qualitative changes in the OBCs of the inflowing gases.  
Thus, Fig. 1 illustrates the OBC for the initial inflowing gas around an accreting object, as well as the forbidden gas region (these energies of the gases are not permissible in any accretion system), which we refer to as the OBC-plane. This OBC-plane is plotted in the $\be - \rob$ plane, where $\rob$ denotes the location of the outer accretion boundary or disc size, usually defined as the boundary where the inflowing gas has an initial inflow velocity $\ve_{\rm ob}\approx 0$, an initial temperature ($\tob$) and an initial specific angular momentum ($\lmdob$). 
$\be$ denotes the total local specific energy parameter of the inflow gas at $\rob$. The $\lmdob$ and $\tob$ are free parameters of the inflowing gases, but the $\lmdob$ has a maximum limit as the Keplerian angular momentum ($\lmk$). The total local specific energy parameter ($B$) encompasses all the energies presented  in the inflowing gas, which includes the kinetic, thermal, gravitational, and rotational energies of the gases in the presented study, and it can be written \citep{nkh97,kg19, k24} as
\begin{equation} B = \frac{\ve^2}{2} + h + \frac{\lambda^2}{2r^2} + \Phi,\label{b.eq}
\end{equation} 
where $h = nc_s^2$ is the specific enthalpy energy of the gas.  
Here, $c_s = \sqrt{\gamma p/\rho}$ is the sound speed, $n = 1/(\gamma - 1)$ is the polytropic index, and $\gamma$ is the adiabatic index. 
The parameter $B$ at $\rob$ can be expressed as
\begin{equation} 
\be = h_{ob} + \frac{\lmdob^2}{2\rob^2} + \Phi,
\label{bob.eq}
\end{equation} 
where $h_{\rm ob}=nc_s^2|_{\rm ob}$ is the specific enthalpy of the gas at $\rob$, assumed the initial inflow velocity ${\ve}_{ob}\sim0$ at $\rob$ and $c_s^2|_{\rm ob} = 2\gamma \thob/\tilde{t}$. Here, $\thob$ is the dimensionless initial temperature of the gas, defined as $\thob=k_{\rm B} \tob/(m_{\rm e} c^2)$, and $\tilde{t}=1+m_{\rm p}/m_{\rm e}$. $k_{\rm B}, m_{\rm e}$ and $m_{\rm p}$ are the Boltzmann constant, mass of electron, and mass of proton, respectively. To analyse the OBC-plane, we assume that the maximum $\thob$ of the inflowing gas might be the virial temperature ($\thvir=k_{\rm B} \tvir/(m_{\rm e} c^2)$). Nonetheless, a central object can accrete gas at any temperature, including super-virial temperatures, if available. The virial temperature is defined as $\tvir=2G\mbh m_{\rm p}/(3k_{\rm B}R)$ \citep{kfm08,k24}. 
 The $\thob$ of the gases is a free parameter, and here, we compared it with $\thvir$.
The $\lmdob$ of the gas is also a free parameter and can vary from $\lmdob<<\lmk$ to a maximum of $\lmk$ at $\rob$.

In panel (a) of Fig. \ref{Fig1}, a solid red line in the left part ($\be<0$) indicates the lowest possible specific energies of the inflowing gases, corresponding to the gravitational potential energy of a particle around the central object. 
Energies lower than the solid red line are not attainable for the gases, and the region on the left side of this line can be defined as a forbidden region. Thus, this solid red line in the $\be<0$ region separates the forbidden and allowed energy regions in the OBC-plane. If we increase any form of energy of the inflowing gases at $\rob$, then $\be$ of the gas shifts only towards the right side of the solid red line in the plane. 
The allowed energy region can again be divided into hot-mode and cold-mode initial inflowing gases, based on the different variations of $\rob$ to $\be$, with both modes separated by the vertical $\be=0$ line. 
A red dotted line in the cold-mode region represents energies $\be\sim\lmk^2/2\rob^2+\Phi$, which have negligible thermal energy $h\sim0$ ($\thob<<\thvir$) compared to gravitational energy. 
The region between the solid and dotted red lines can represent the inflowing gases with $\lmdob<\lmk$ in the cold-mode region. A hot-mode region is divided into super- and sub-virial regions by the solid violet line (if inflowing gas has $\lmdob=0$) or dotted violet lines (if $\lmdob=\lmk$). The region on the right side of the dotted line in the hot mode is called the super-virial region. We believe that the allowed energy region with $\rob>10^3\rg$ can illustrate the accretion boundaries in the BXBs and AGNs. The allowed energy region with $\rob\lsim10^2\rg$ can denote the accretion boundaries in the TDEs, which have disc sizes smaller than the AGNs. Thus, these OBCs can provide a unified picture of the accretion boundaries around various accreting black holes, which we refer to as the general "OBC-plane." This OBC-plane is also equally applicable to the accretion boundaries in other accreting sources, such as protostars (YSOs).

Furthermore, the inflowing gases at the $\rob$ may have some initial inflow velocity, implying that ${\ve}_{\rm ob}\neq0$ means not negligible compared to other local velocities, and yet $\be$ will remain in the OBC-plane. For example, the winds of the companion star in BXBs may carry an initial injection velocity if they have not lost their radial velocity upon colliding with the disc or accretion boundary. It is also plausible that the winds of the companion star can lose their partial or full angular momentum when they collide with each other or impact the disc or accretion boundary. The key point is that the inflowing gases at the accretion boundary can exhibit numerous possibilities and combinations, such as cold or hot mode gas with sub-Keplerian angular momentum, in addition to gas with some initial inflow velocity, so we included two cases, V and VI, in this study from both cold and hot mode regions.
\subsection{Simulation input boundary conditions}\label{subsec:SBC}
The OBCs for the accretion flows are illustrated in Fig. \ref{Fig1}a (OBC-plane), which displays the initial specific energies of the inflowing gases at the outer boundary radial distances from the central objects. The initial input parameters for our simulation at the simulation boundary can be directly obtained from the OBC-plane. 
Implementation of the proper initial boundary conditions is crucial when studying the accretion flow around the accretors, such as BH XRBs and AGNs. These sources have a much larger scale length compared to current simulation studies. Applying initial boundary conditions that reflect the realistic size of the disc in simulations is very challenging and time-consuming. Interestingly, semi-analytical studies have shown that the formation of various types of advective flows is generally independent of $\rob$. The different semi-analytical solutions can be generated for discs of any size, except for steady shock solutions in the accretion flow, which can occur when $\rob > 10^3\rg$ \citep{ky21}. Our primary goal is to generate the various types of advective flows as found in the semi-analytical studies. So it would be suitable to use initial input parameters of the simulation from the semi-analytical solutions, which can provide opportunity to test the code and a comparative study with the semi-analytical studies, including predictions of the semi-analytical results, especially, nature of the accretion solutions and corresponding connections of the OBCs in the OBC-plane, shock formation and location, and independence of the nature of the solutions with $\rob$. 
In this approach, the first step is to compute the accretion solutions using the steady-state accretion model (refer to \cite{kc13} and \cite{r25} for details on the basic equations and numerical methods). {However, for the convenience of the reader, we have briefly described the basic equations in the Appendix \ref{sec:tsss}. The steady-state or semi-analytical solutions are computed by using OBCs from the OBC-plane at a $\rob$.} 
The flow variables of these semi-analytical solutions at a radius (which depends on the simulation box size) serve as the initial input parameters/conditions for the simulation, which can be called simulation BCs.

The $\rob$ of the sources can be very large, making it challenging to simulate results with extensive boundaries while maintaining the appropriate resolution. So, we defined another boundary as the simulation boundary or injection radius ($\rinj$). Computing semi-analytical solutions with a large boundary is comparably easier by allowing the theoretical boundary to represent the actual boundary of the sources. 
Interestingly, we have found that the types of solutions are independent of the $\rob$ \citep{ky21,k24}, which can be generated for any disc size, except for the standing shock solutions. The standing shock solution requires a large $\rob>10^3\rg$ in the steady-state accretion models \citep[details in][]{ky21}. 
So, we have chosen two different outer boundaries, which are $\rob=400\rg$, and $2400\rg$. Choosing these values of the $\rob$ has two main reasons, first reason, we want the outer critical point (or sonic point) should form within $50\rg$ from the BH so we can used sub-sonic initial boundary conditions for simulation, unlike previous studies have used super-sonic initial conditions \citep{msc96,lmc98,lck16,gbc20,soa21,g25} and location of the outer critical point in the accretion flow depends on the $\rob$ and $\lmdob$ \citep{kc13,ky21,r25}. Second reason, the range of $\rob$ can be divided into two parts based on the formation (or no formation) of the standing shocks in the semi-analytical studies, which are $\rob\lessgtr 10^3\rg$ range. 
The $\rob=400\rg$ can generate all types of accretion solutions, except the standing shock solution. So we used another $\rob=2400\rg$  
can generate all types of solutions, including standing shock in the semi-analytical model. 
Doing so, we can predict the robustness of the nature of accretion flows in both semi-analytical and simulation studies.

If we draw a horizontal line at a particular $\rob$ in the OBC-plane, then this line can pass through three allowed regions: sup-virial cold-mode, sub-virial hot mode, and super-virial hot-mode. Interestingly, the sub-virial regions of the OBC-plane always give two types of accretion solutions, which are single sonic point smooth solutions or multi-transonic solutions in the semi-analytical study \citep{ky21,k24}. However, the super-virial region can give two types of solutions or only a smooth solution, depending on the $\rob$. The multi-transonic solutions can have a standing shock in the semi-analytical accretion solutions. 
The panel (b) of Fig. \ref{Fig1} illustrates the initial energies of the inflowing gases in $\be - c_s|_{ob}$ plane, which is plotted corresponding to the horizontal cross-section in the OBC-plane at $\rob=400\rg$. The initial energies of the inflowing gases for $\rob=400\rg$ are constrained between both lines (solid and dotted). It can be considered a zoomed-in version of the OBC-plane for a particular $\rob$. Here, $c_s|_{ob}$ denotes the sound speed of the inflowing gases and is plotted up to the sound speed corresponding to the virial temperature. The region $\be<0$ represents the OBCs of the cold-mode gas, and $\be>0$ represents the OBC of hot-mode gas. We chose four points/locations, which are represented by $\blacksquare$ symbols in the panel (b) of Fig. \ref{Fig1}. Each point is chosen with the sub-virial temperature at $\rob$ and denoted from case III to case VI.
The OBCs corresponding to these four symbols $\blacksquare$ yield different types of typical accretion solutions, which can exhibit single or multiple critical points as predicted in many studies \citep[][]{c89,ac90,lgy99,cd07,bdl08,kscc13,kc14,ck16,kc17,ky21}. The super-virial region for $\rob=400\rg$ has only a single sonic point solution. However, the super-virial region at $\rob=2400\rg$ has both single and multi-sonic point solutions.

We have selected six typical cases, depending on the three different regions (sup-virial cold-mode, sub-virial hot-mode, and super-virial hot-mode) of the OBC-plane and the nature of the solutions. 
First, two cases are selected from the super-virial region of the OBC-plane, and the corresponding OBC points are denoted by the symbols $\blacksquare$ and $\bullet$ in that region and named as cases I and II. The other four cases are taken from the sub-virial regions as four OBC points denoted in the right panel of Fig. \ref{Fig1}.
We generated steady-state accretion solutions from these points of OBCs with the help of equations in the Appendix \ref{sec:tsss}, and corresponding simulation results
 are presented in the result section \ref{sec:result}. We used other basic parameters, an adiabatic index $\gamma=1.4$ and a viscosity parameter $\alpha=0.01$ for all cases in both semi-analytical and simulation results. The semi-analytical solutions are obtained by solving three differential equations (\ref{ssdv.eq}-\ref{ssdcs.eq}) simultaneously. The details of numerical methodology for solving those equations are given in \cite{kc13}. The flow variables of some semi-analytical solutions with radial distances are plotted with a solid line in the right panels of Figs. \ref{Fig3} and \ref{Fig8R}, here, these solutions are plotted up to the $\rinj$. For an overview, the examples of the full solutions up to $\rob$ can be found in the previous studies \citep{kc13,kg19,ky21,r25}. Since each region of the OBC-plane gives a smooth and shocked accretion flow. Therefore, we named each case in this study in order to differentiate them, which is based on the nature (smooth or shocked) of the accretion flow in the simulation results and the corresponding location of the OBC point in the OBC-plane. For example, Case I has smooth flow and is generated from the OBC point in the super-virial region of hot-mode ($\be>0$) gas in the OBC-plane, so we named it smooth (super-hot) flow; Case II exhibits shock in the flow and generated from the OBCs in super-virial region, so we named shocked (super-hot), Case III has smooth flow and is generated from the OBC point in the sub-virial region of cold-mode gas in the OBC-plane, so we named it smooth-cold flow, 
 and similarly others are named. 

\section{Results}\label{sec:result}
As noted earlier, our simulation code requires the initial input parameters in the form of flow variables at the simulation boundary ($\rinj$). 
The main initial input/injection parameters include the local rotational velocity ($v_{{\phi}_{inj}}=\lmdinj/\rinj$), the local velocity of inflowing gas ($\vinj$), and the local sound speed ($\csinj$) at the ($\rinj$). Other injection parameters, including local gas density ($\rhinj=\dot{M}/(4\pi\vinj\rinj\hinj)$) and local disc height ($\hinj=(2\rinj/\gamma)^{1/2}\csinj(\rinj-1)$) at $\rinj$, are derived from the main injection parameters by using the mass conservation equation (\ref{mdot.eq}) and the vertical hydrostatic equilibrium condition (\ref{hh.eq}), respectively. 
Consequently, we have mentioned the values of the main injection (or initial input) parameters for the generation of the simulation results in the caption of each figure. 
The main injection parameters ($\lmdinj=\lambda, \vinj=\ve, \csinj=\cs$) are obtained from solving three differential equations in Appendix \ref{sec:tsss}.
The $\rhinj$ is calculated in each result with $\dot{M}=0.01\medd$, where  $\medd=1.44\times10^{18}\mbh/\sun~g/s$ is the Eddington accretion rate { when assumed $\eta=0.1$ in $L_{Edd}=\eta\dot{M}_{Edd}c^2$} and used $\mbh=10\sun$. 
Here, we have presented the six cases I to VI (denoted in Fig. \ref{Fig1}) to represent three typical regions (cold-mode, sub-virial hot mode, and super-virial hot-mode) of the OBC-plane, with each region corresponding to a specific class of accretion solutions as identified in the semi-analytical studies.

The semi-analytical solutions are generated corresponding to the six cases, and the simulation initial input parameters at the simulation boundary $\rinj$ are obtained from them. The initial setup of the simulation box at the time of $t = 0.0$ seems to show a vacuum, but there exists a background density, which is about $10^3 - 10^5$ times lower than the inflow density and depends on the property of the inflow. The matter with the initial input parameters is injected through the right-hand side boundary wall (or the ghost cell) in the simulation box. The resolutions and size of cells are uniformly distributed in the simulation box.
The scale height and matter density of the flow at the simulation boundary are also determined by the steady-state model equations (Appendix \ref{sec:tsss}).  To achieve better computational results with a sink size of $1.0\rg$, more cells and computational time are needed. Therefore, black hole simulations usually employ sinks ranging from $1.5\rg$ to $3.0\rg$. For convenience, we used a sink radius of $1.5\rg$ in most cases, which produces suitable results at lower resolutions than a sink radius of $1.0\rg$, while both produce the same disc structures. Fig. \ref{Fig8R} in this paper is simulated using a sink radius of $1.0\rg$ with higher resolution than the other cases (except Fig. \ref{FigA} in Appendix \ref{sec:hr}), because the higher resolution provides better accuracy for the shock location when compared with the semi-analytical solutions. The resolutions and sizes of the simulations are mentioned in the description of each figure. All the simulated results are plotted here when the accretion flows achieved a steady or quasi-steady nature in the simulation box.
\subsection{Comparison of viscous simulation results with semi-analytical solutions}\label{subsec:Bondi}
In this subsection, we demonstrated the consistency between the simulation and semi-analytical results.
Firstly, we investigated two popular semi-analytical solutions for the viscous flow: Bondi-type smooth and shocked accretion flows, which are obtained for OBCs with $\lmdob=\lmk$ and super-virial temperatures ($\thob>\thvir$) of the inflowing gases at the outer boundary. However, the temperature distribution of these flows has a sub-virial temperature distribution across the flow except for the outer part of the disc \citep{k24}. Many authors have studied these types of global accretion solutions (shocked and Bondi-type) through semi-analytical as well as simulations approaches \citep{c89,ac90,ct95,msc96,c96,lgy99,y99,dc99,cm06,cd07,bdl08,lrc11,kc13,kc14,kcm14,kc17,adn17,ck16,ddp20,ky21,smd22,md24,dcj24,r25}. The OBCs in the super-virial region consistently produce saddle (`X') type outer critical points, which yield global accretion solutions crucial for the semi-analytical studies. The super-virial temperature of the inflowing gases at the outer boundary remains a contentious topic, as it results in a notably high temperature \citep{k24}. We employed these OBCs (in the super-virial region of the OBC-plane) to investigate the accretion flows, which provide two types of accretion solutions: smooth and shock flow. So, we have chosen two OBC points that give both types of solutions. 
The locations of these OBC points are indicated in the OBC-plane of Fig \ref{Fig1} using symbols \( \blacksquare \) and $\bullet$ in the super-virial region. 
Firstly, we generated the semi-analytical solutions for each OBC point. Then, we picked the injection parameter from those semi-analytical solutions at $\rinj$ and generated the simulation results. We presented both results and compared them in the following paragraphs.

\begin{figure}
\includegraphics[angle=0, width=0.45\textwidth]{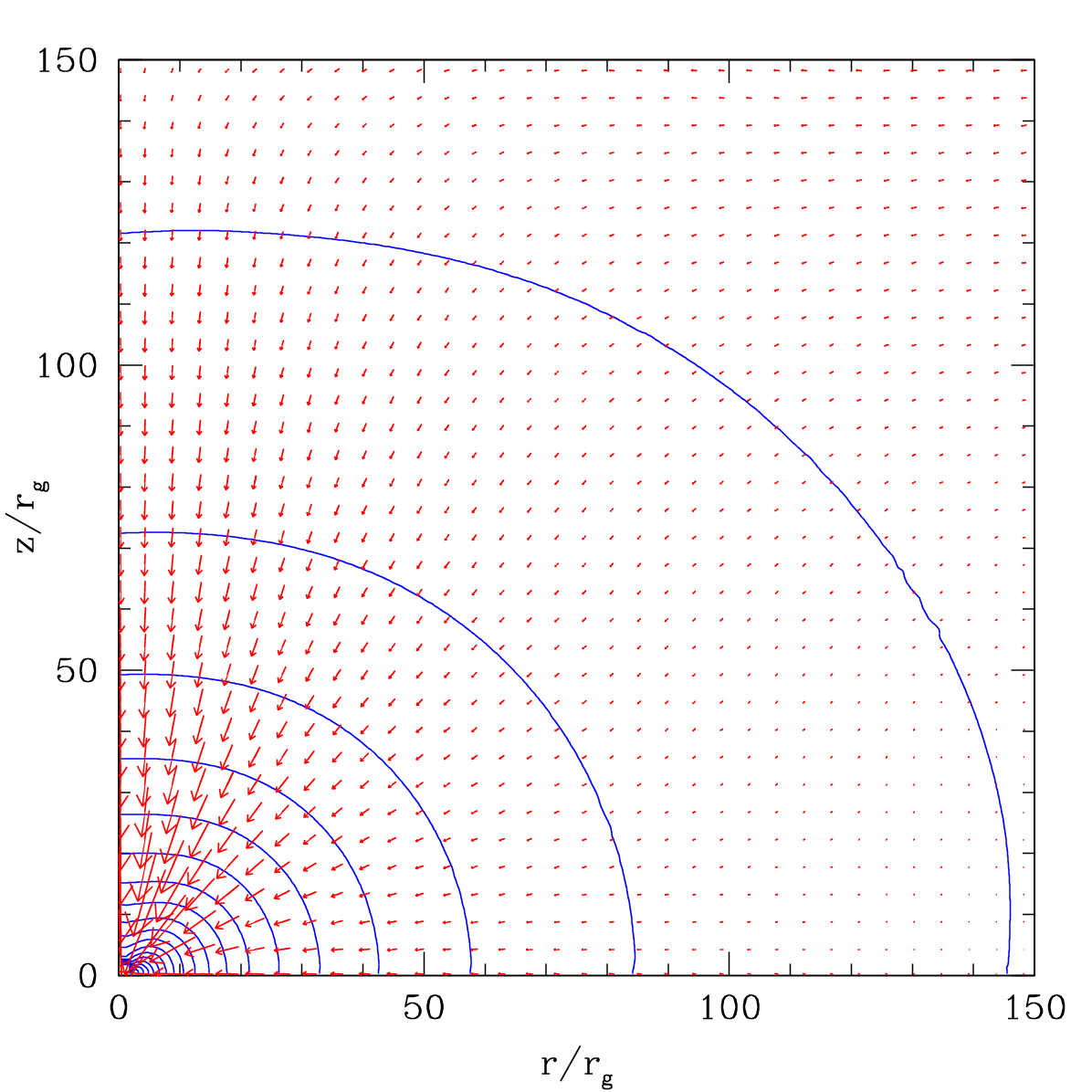}
\includegraphics[angle=0, width=0.45\textwidth]{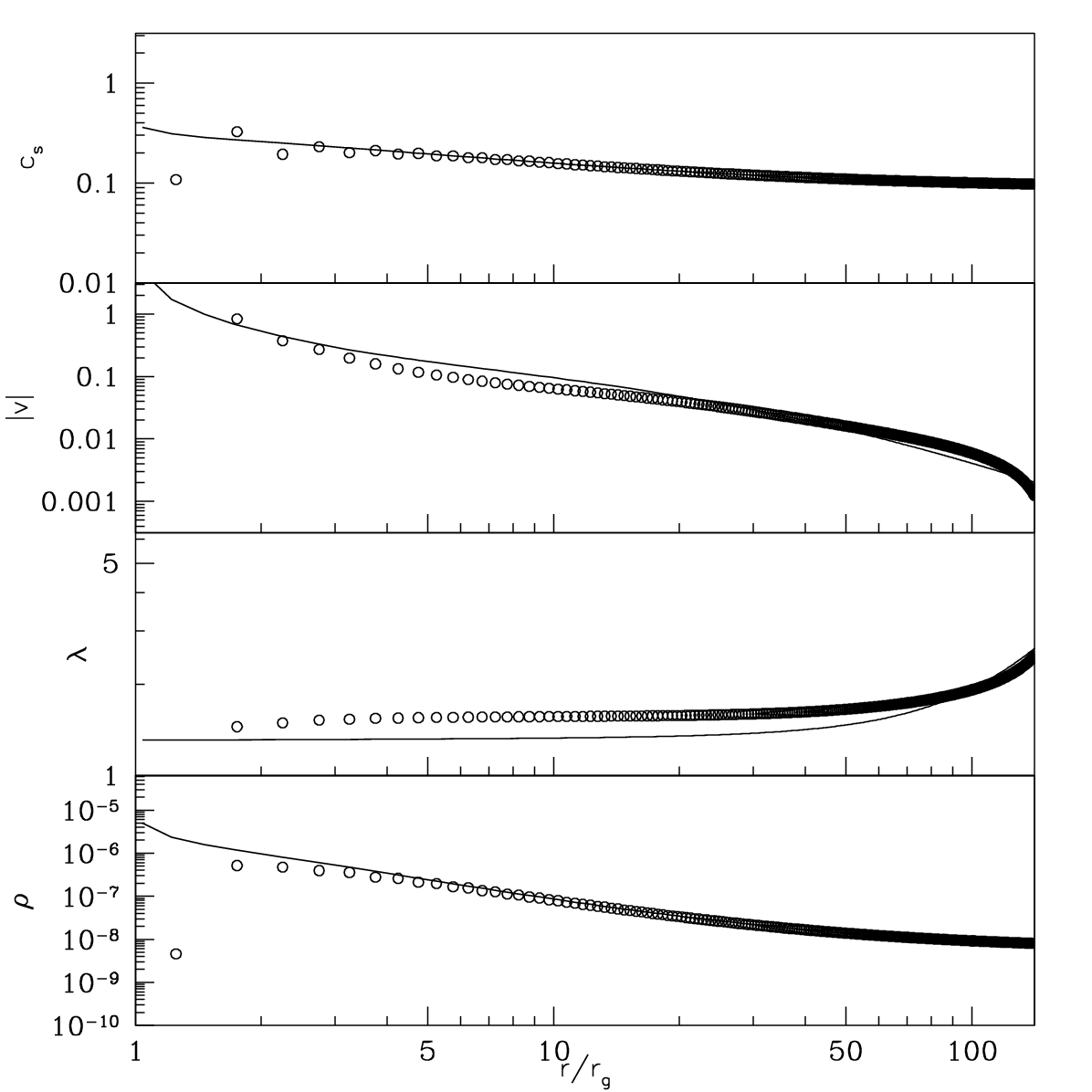}
\caption{The density contours and velocity vector field in the $r-z$ plane for the smooth (super-hot) flow or Bondi-type spherical flow in the left panel. The right panel compares flow variables of the semi-analytical (solid line) and simulation (open circles) results. {Simulation input parameters are $\csinj=9.666187\times10^{-2}c, \vinj=-1.779581\times10^{-3}c; \lmdinj=2.814507\rg c$, and $ \rhinj=7.836585\times10^{-9}~g/cm^3$ at $\rinj=150\rg$. Here, eighteen density contours are labelled in log scale with the increment of $0.1$ from the boundary to the centre in density range, $8.0\times10^{-9}$ to $4.001\times10^{-7} ~g/cm^3$.}
\label{Fig3}}
\end{figure}
Case I gives a smooth accretion flow and is plotted in Fig. \ref{Fig3}, when the flow becomes steady after some simulation time. We used a resolution of $300\times500$ cells in a box size of $150\rg\times250\rg$. 
The contours and velocity vectors indicate density variations and flow directions, with velocity magnitudes ($|\mathbf{v}|=\sqrt{\vr^2+v_z^2}$) in the panel, respectively. Here, $\mathbf{v}=\mathbf{v_r}+\mathbf{v_z}$ is the velocity vector of the flow. The density contours are labeled in log scale levels increasing with the increment $0.1$ towards BH from the simulation boundary and decreasing in the $z$-direction from the equatorial plane.
This flow achieved a quasi-spherical structure of accretion flow, which can represent the Bondi-type spherical flow.
This type of flow has the lowest $\lambda$ distribution (Fig.\ref{Fig6}) among the advective flows and exhibits no outflows. The lowest $\lambda$ distribution is also confirmed in the semi-analytical studies \citep[references therein]{k24}. The left panel in Fig. \ref{Fig3} displays variations of the density contours (solid blue curves) and velocity vectors (red arrows), indicating the direction of motion of the accretion flow. In this scenario, the simulation box is filled with moving inflow gas, all of which converges towards the BH, showing no signs of outflows. 
The right side of Fig. \ref{Fig3} illustrates variations of flow variables at the equatorial plane for simulated (open-circle curve) and semi-analytical Bondi-type (solid curve) accretion flows as a function of radial distance ($r$). In these panels, the variations in sound speed ($c_s$), inflow velocity (${v}$), specific angular momentum ($\lambda$), and flow density ($\rho$) are presented from the top to bottom panels, respectively. Notably, both simulation and semi-analytical flow variables are closely aligned, demonstrating the robustness of the solutions and the codes. The flow variables, $c_s, v$, and $\rho$ are increasing towards BH. However, $\lambda$ is decreasing towards BH due to the outward transportation of it.

\begin{figure}
\centering
\includegraphics[angle=0, width=0.48\textwidth]{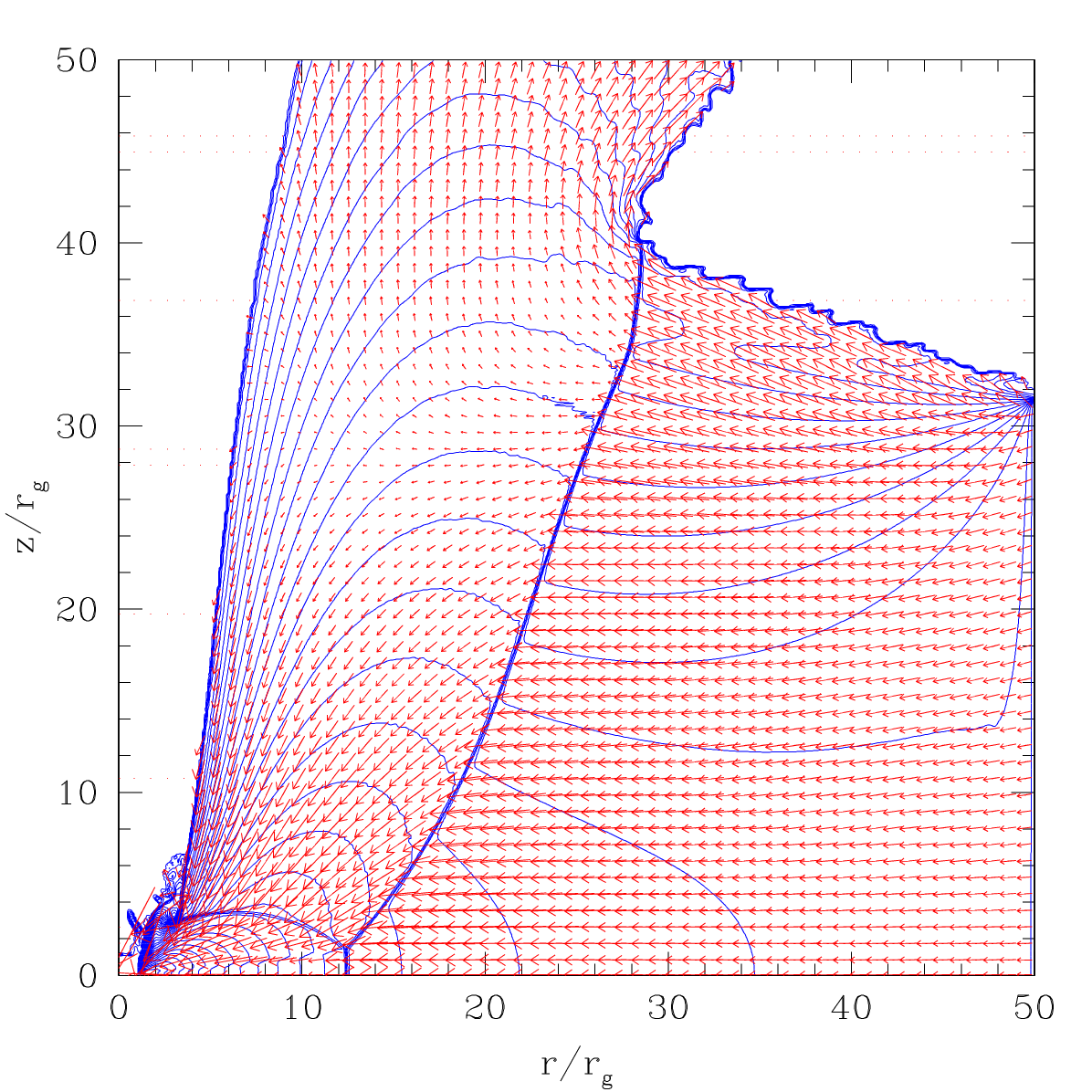}
\includegraphics[angle=0, width=0.45\textwidth]{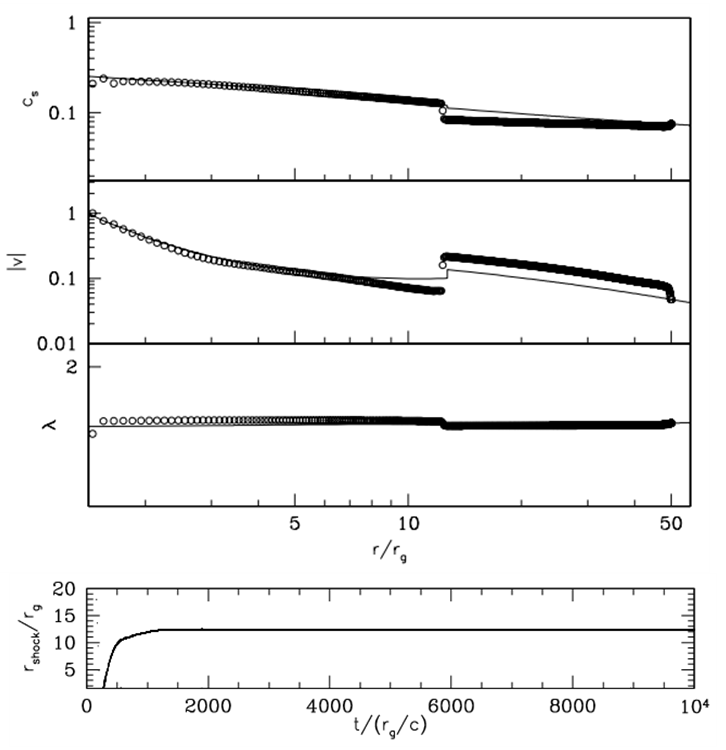}
\caption{Structure of shocked disc with inflow-outflow regions is presented with density contours and velocity vectors in the $r-z$ plane. The post-shock region has the outflows. The right panel shows the simulation flow variables (open circles) and semi-analytical results (solid lines). The bottom panel shows the variation of the shock location on the equatorial plane with simulation time. Simulation input parameters are $\csinj=7.513592\times10^{-2}c, \vinj=-4.732474\times10^{-2}c; \lmdinj=1.526311\rg c$ at $\rinj=50\rg$. 
\label{Fig8R}}
\end{figure}
The disc structure depicted in Fig. \ref{Fig8R} shows shock in the flow and is simulated from the OBC, which is denoted by the $\bullet$ symbol at $\rob=2400\rg$ in the super-virial region of the OBC-plane, hence named it shocked (super-hot) flow. In this case, we used resolution {$600\times600$} cells in a $50\rg\times50\rg$ box, which produces a shock at the location, which is very close to the semi-analytical solution shock at $12.69\rg$ in the flow.
This case has outflows in the post-shock region, and the outflow velocity reaches $\sim0.1c$ at $z\sim50\rg$. Here, the outflows in the post-shock region are generated due to the extra thermal pressure in the flow, which is generated promptly by conversion of the inflow kinetic energy into thermal energy at the shock location.  
We have plotted the flow variables from the simulation results (open circle) and compared them with the semi-analytical solution (solid line) on the right side of Fig. \ref{Fig8R}. The panels on the right side of Fig. \ref{Fig8R} illustrate the variations of $c_s, v$, and $\lambda$ from the top to the second bottom panels, respectively. The shock location and the variations in flow variables from the simulation closely align with the semi-analytical results. At the shock location, the flow variables are changed promptly, which shows a shock transition. The $v$ is sharply decreased, and consequently $c_s$ is sharply increased, which increases thermal pressure in the post-shock region. This sharp increment can generate outflows along the $z-$ direction. The bottom panel shows the variation of the shock location with simulation time. The shock location becomes steady after a certain amount of time.
These findings again demonstrate the robustness of the simulation by following the semi-analytical solutions. 
We have generated more disc structures based on the predictions of the semi-analytical solutions with their OBCs locations in the OBC-plane.
We presented additional simulation results with various OBCs in the following subsections.
\subsection{Structures of accretion flows with various sub-virial OBCs}\label{subsec:vobcs}
\begin{figure*}
\centering
\begin{subfigure}{.45\textwidth}
		\includegraphics[width=\textwidth]{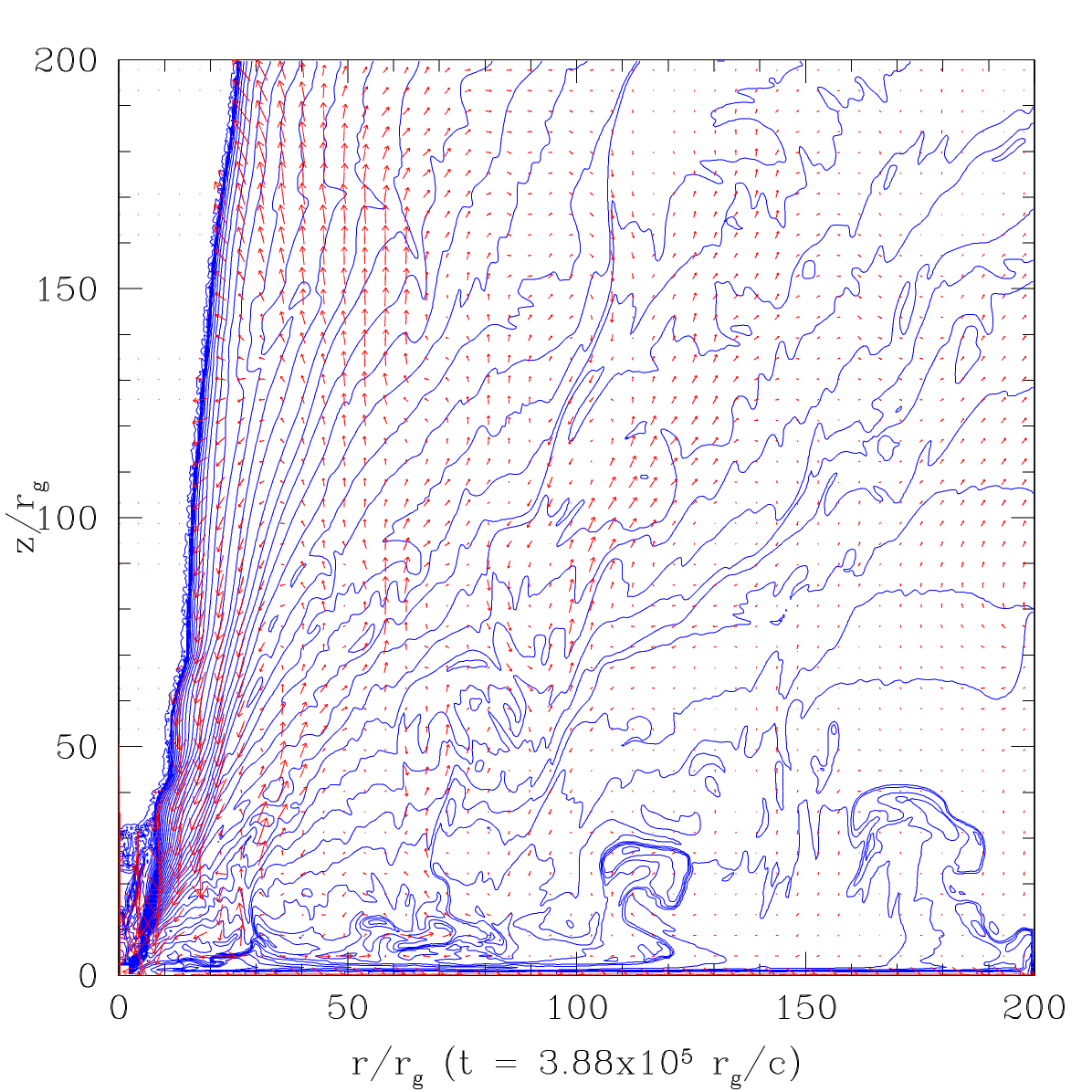}
		\caption{Case III - Smooth-cold}
	\end{subfigure}
	\begin{subfigure}{.45\textwidth}
		\includegraphics[width=\textwidth]{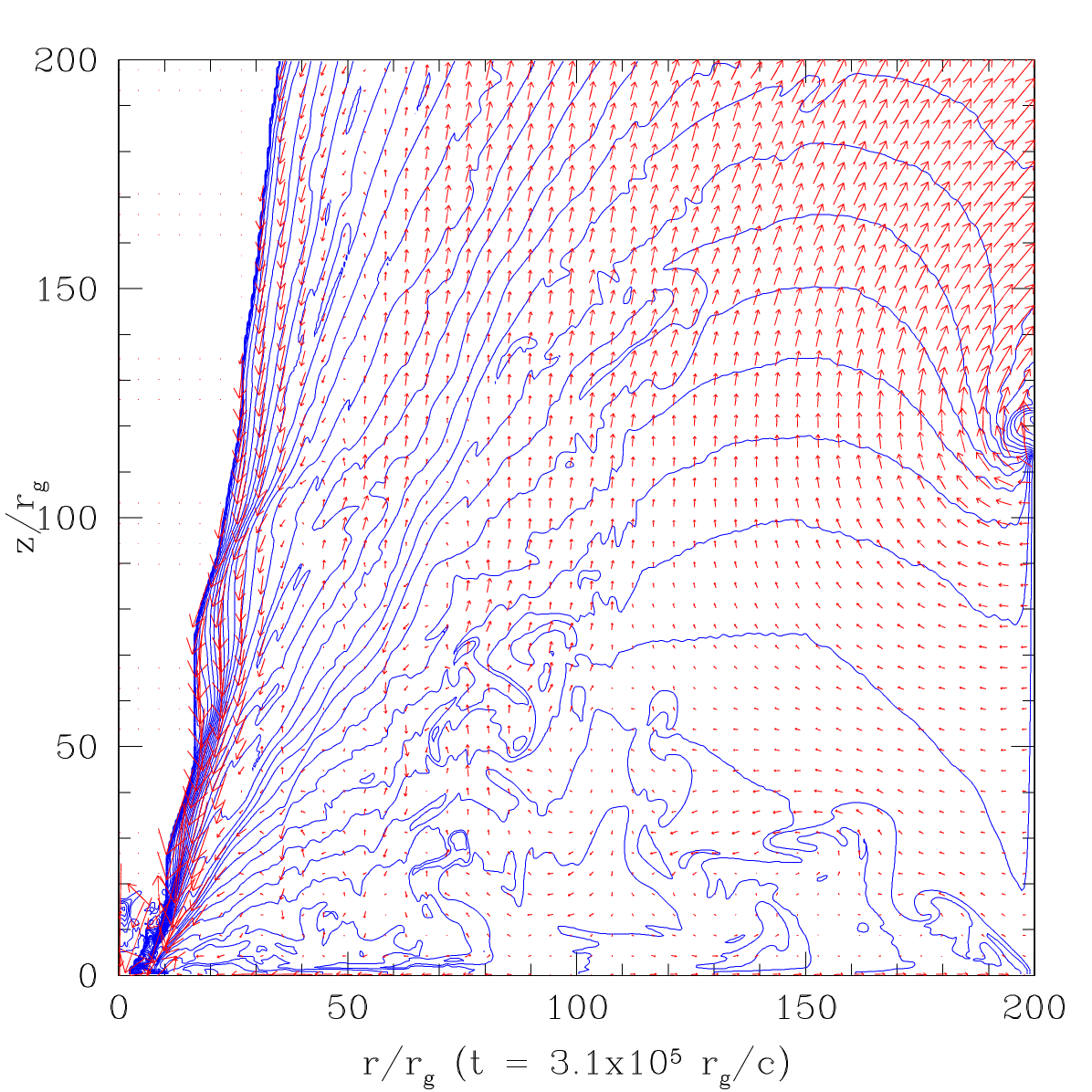}
		\caption{Case IV - Smooth-hot}
	\end{subfigure}
	\begin{subfigure}{.45\textwidth}
		\includegraphics[width=\textwidth]{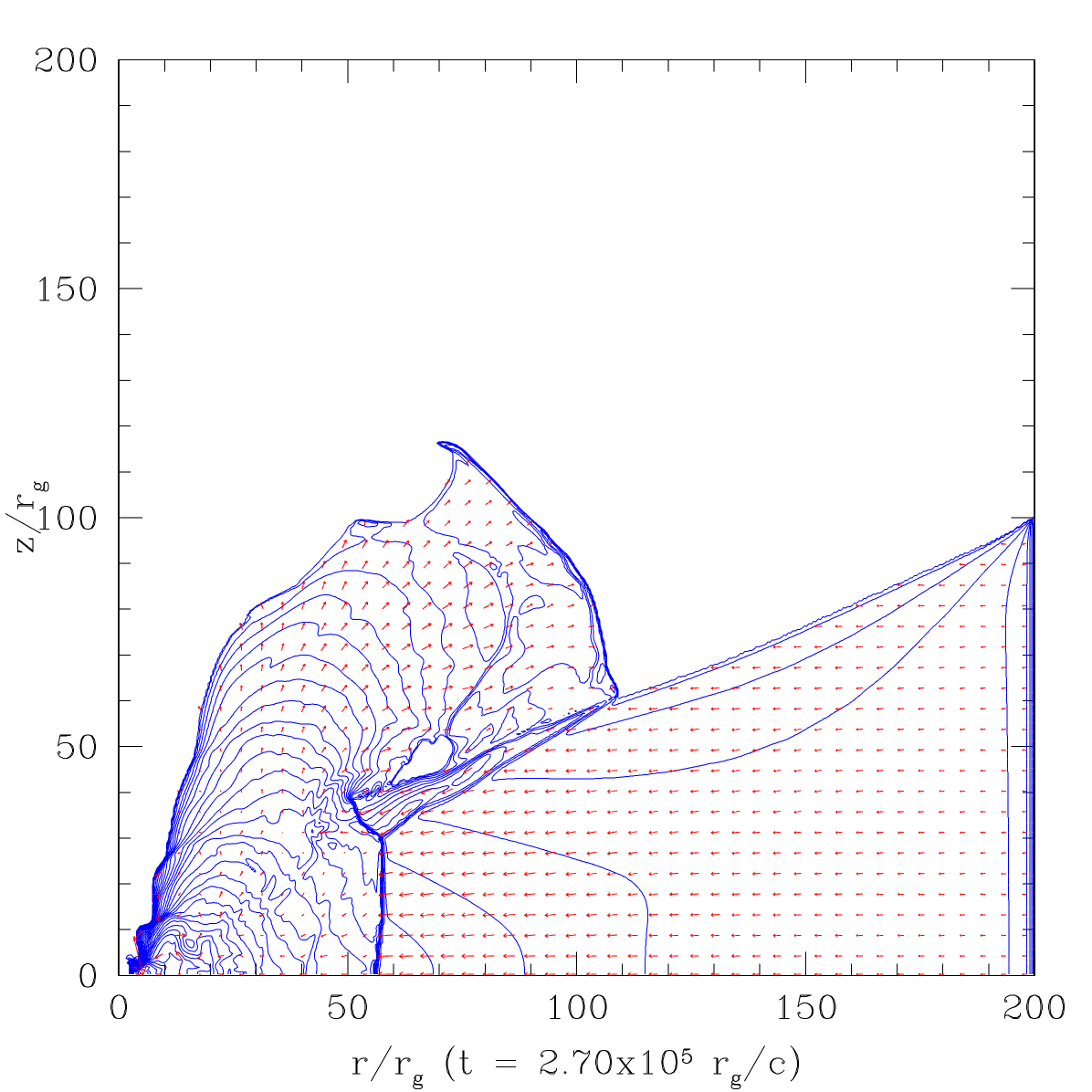}
		\caption{Case V - Shocked-cold}
	\end{subfigure}
	\begin{subfigure}{.45\textwidth}
		\includegraphics[width=\textwidth]{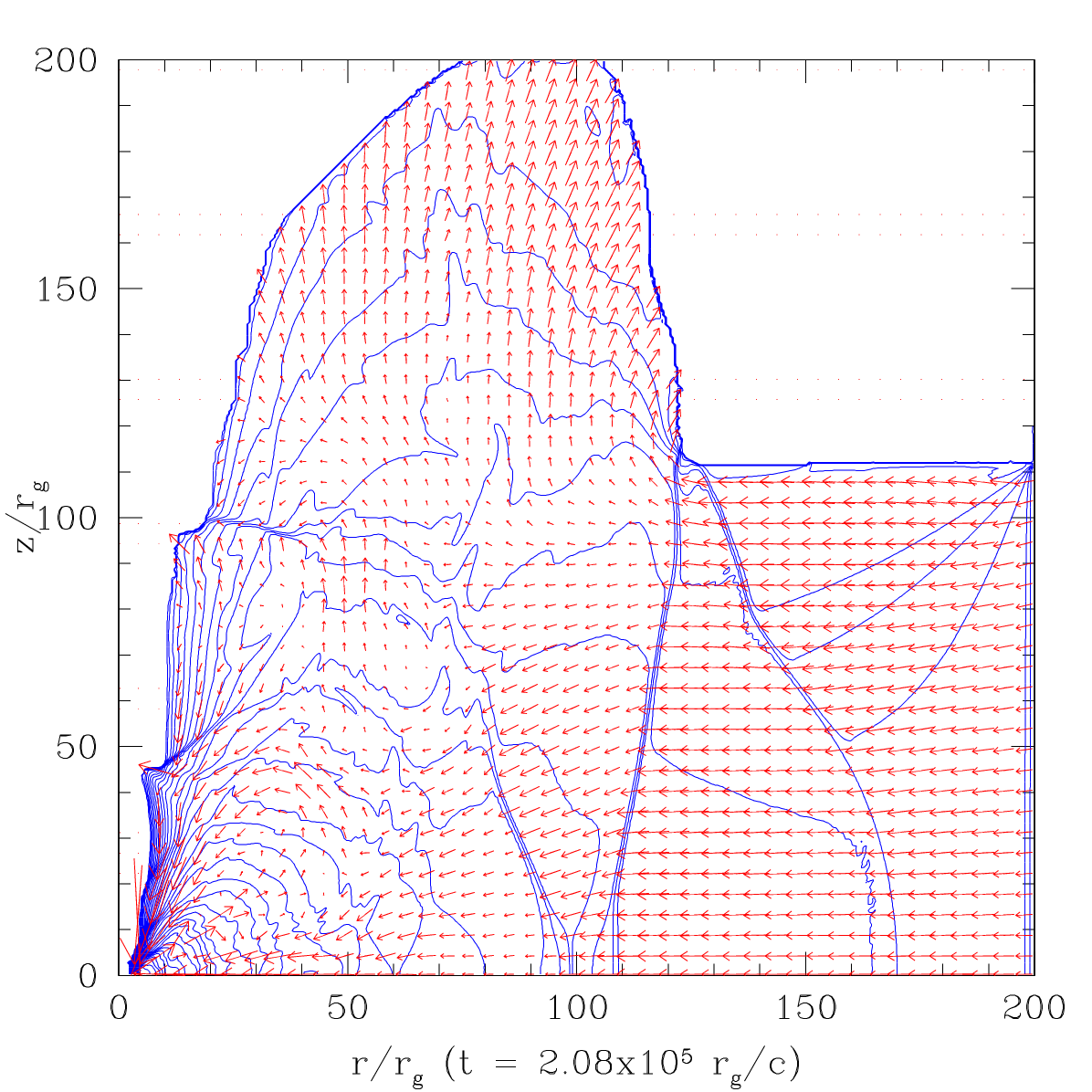}
		\caption{Case VI - Shocked-hot}
	\end{subfigure}
\caption{Variation of density contours and velocity field in $r-z$ plane. These images represent different accretion flows with different OBCs and name them according to their nature and locations of the OBCs in the OBC-plane, which are named smooth-cold (panel a), shocked-cold (panel b), smooth-hot (panel c), 
and shocked-hot (panel d) types of accretion flows. {Panel (a) has the input parameters at  $\rinj=200\rg$, the $\csinj=2.323226\times10^{-2}c, \vinj=-1.234235\times10^{-4}c; \lmdinj=8.611521\rg c$, 
panel (b) has $\csinj=3.334912\times10^{-2}c, \vinj=-2.682483\times10^{-4}c; \lmdinj=6.696894\rg c$. 
panel (c) has $\csinj=2.932613\times10^{-2}c, \vinj=-4.368247\times10^{-3}c; \lmdinj=1.891245\rg c$, 
and panel (d) has $\csinj=3.270309\times10^{-2}c, \vinj=-1.458542\times10^{-2}c; \lmdinj=1.786880\rg c$. 
The density contours are levelled in log scale with an increment of $0.1$, for instance $\rhinj=1.376054\times10^{-9}~g/cm^3$ in panel (d), which increases towards the BH centre, and decreases along the $z$-direction from the equatorial plane.}
\label{Fig4}}
\end{figure*}

As noted earlier, the OBCs with sub-virial temperature have divided into two regions in the OBC-plane, which are the cold-mode region and the sub-virial hot-mode region. Each region gives two types of accretion solutions.
So, we have selected four cases from the sub-virial region of the OBC-plane based on the typical nature of the semi-analytical accretion solutions, and the OBC points are located in panel (b) of Fig. \ref{Fig1}, denoted by symbols $\blacksquare$ points and represent cases III to IV. 
The simulations corresponding to these OBC points are generated and presented in Fig. \ref{Fig4}, and each case is named according to the location of the OBC points in the OBC-plane and the nature of the solutions.
We mentioned the simulation parameters corresponding to each simulation disc structure in the caption of Fig. \ref{Fig4}.  
The two disc structures of the first row in Fig. \ref{Fig4} are generated from the simulation parameters, which are taken from the smooth semi-analytical solutions. Both smooth semi-analytical solutions are generated for cases III and IV. 
Both cases have the same $\lmdob=\lmk$ of initial inflowing gases, but different $\thob$ from the cold-mode ($\be<0$) and hot-mode ($\be>0$) regions. 
However, the two simulation disc structures of the second row in Fig. \ref{Fig4} are generated from the simulation parameters that are taken from the semi-analytical solutions with the OBC points (cases V and VI in Fig. \ref{Fig1}b). Here, both OBC points represent initial inflowing gases with $\lmdob<\lmk$ at the boundary from the cold-mode and hot-mode regions. 
Both semi-analytical solutions corresponding to these OBCs have two sonic points with wholly sub-Keplerian angular momentum distributions. The OBCs of these two cases are based on the possibility that the inflowing gas can have sub-Keplerian angular momentum at $\rob$, as discussed earlier in the last paragraph of section \ref{sec:obcs}, before sub-section \ref{subsec:SBC}.
The panels in Fig. \ref{Fig4} illustrate the structures of accretion flows in the $r-z$ plane with the resolution of $400\times400$ cells in a $200\rg\times200\rg$ box size. The contours and velocity vectors indicate density variations and flow directions ($\mathbf{v}/|\mathbf{v}|$) with velocity magnitude ($|\mathbf{v}|$) in each panel, respectively.  
The density contours are labeled in log scale levels increasing with the increment $0.1$ towards BH from the simulation boundary and decreasing in the $z$-direction from the equatorial plane.

\begin{figure}
\includegraphics[width=0.8\textwidth]{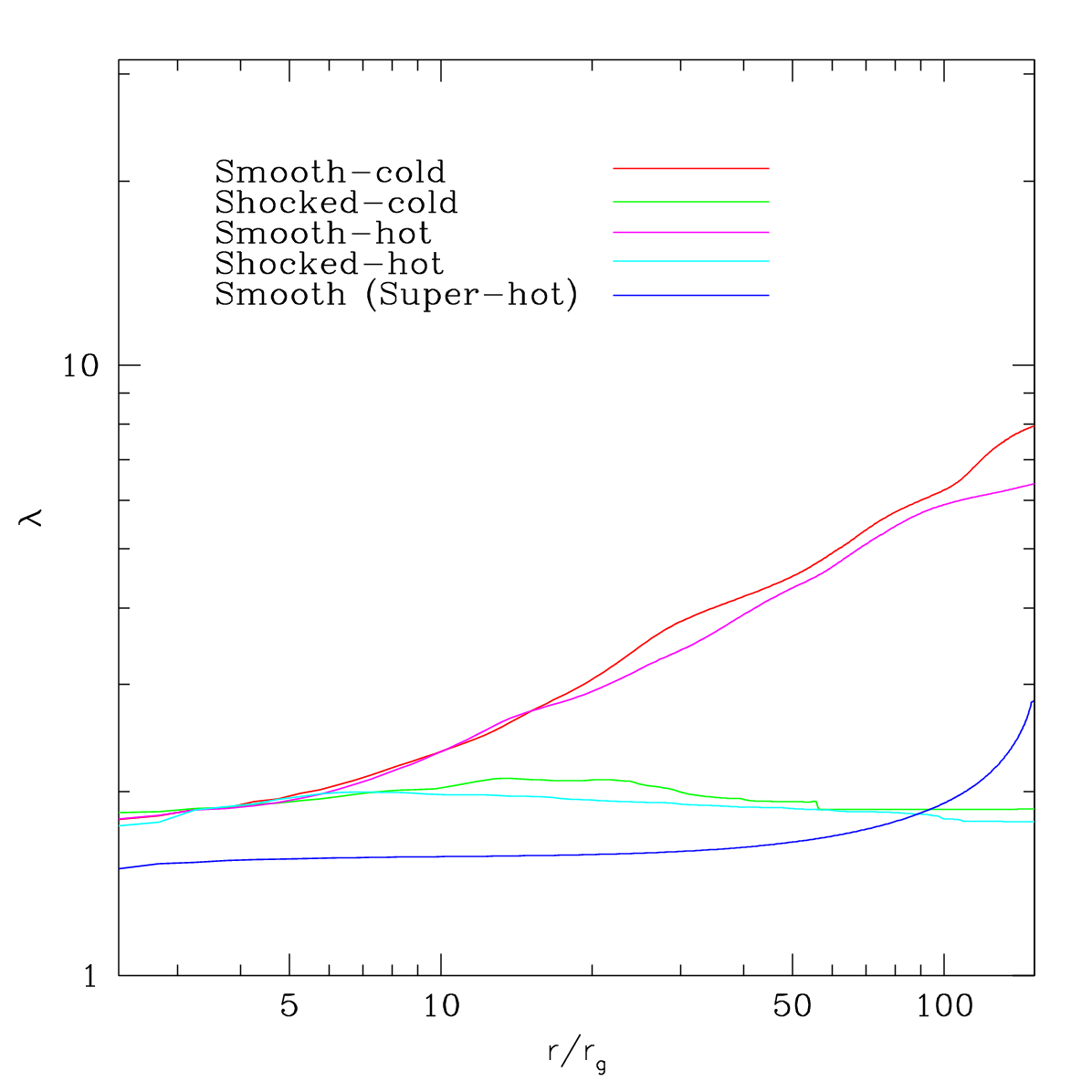}
\caption{Variations of the specific angular momentum $(\lambda)$ with radial distance ($r$) of various types of accretion flows. Each curve represents a different accretion solution, and their names with line colour are mentioned in the figure. All the solutions are generated from the same $\rob=400\rg$ but different $\thob$ or $\lmdob$. Here, the smooth-cold and the smooth (super-hot) flows are generated from the lowest and highest $\thob$ of the inflowing gases, respectively. The shocked-cold and shocked-hot flows are generated with sub-Keplerian $\lmdob$.
\label{Fig6}}
\end{figure}

Case III in panel (\ref{Fig4}a) presents a smooth or shock-free flow, generated using the injection parameters at $\rinj$, which are mentioned in the caption of the figure. 
This simulation flow is named the smooth-cold flow based on the nature of the solution and location of the OBCs in the OBC-plane, 
which follows the ADAF-type or shock-free flow of semi-analytical studies \citep{nkh97,lgy99}. This case exhibits weak outflows. 
Similarly, Case IV in panel \ref{Fig4}b represents the smooth accretion flow, termed smooth-hot, that produces outflows or winds. 
This flow is generated from the input parameters, which are computed from the OBCs in the hot-mode region and follow the ADAF-thick flows in the semi-analytical studies \citep{lgy99,ky21,k24}. This smooth-hot flow produces stronger outflows, which can be due to its higher thermal energy and lower angular momentum distribution  (see Fig. \ref{Fig6}) compared to the smooth-cold flow in panel \ref{Fig4}a. This can be easily seen by the directions and magnitude of the velocity vectors along the $z-$ direction in both \ref{Fig4}a and \ref{Fig4}b panels. 
Interestingly, both disc structures have the outflow in the whole disc of the simulation box, which is similar to past simulation studies \citep{ybw12,ywb12,yyo14} as well as 2D semi-analytical studies \citep[][references therein]{kg18}. Here, the outflows are thermally generated with the support of centrifugal forces.
The centrifugal force has two components in the disc: a tangential component and a radial component. The tangential component acts toward the equatorial plane, allowing the formation of a disc around that plane, while its effect can be suppressed or balanced by the thermal force, leading to the formation of either geometrically thin or thick discs or quasi-spherical flow. However, the radial component of the centrifugal force with the local thermal pressure opposes gravity. Thus, the combined effects of thermal and centrifugal forces within the disc can drive outflows in the HD regime if both forces successfully counteract gravity. 
Relying solely on either centrifugal or thermal forces may not generate outflows, which can be due to all the forces generated through accretion processes in the disc being connected to each other.   
For instance, in two extreme cases—a thin Keplerian disc (high $\lambda$ but weak thermal pressure) and a Bondi-type spherical flow (high thermal pressure but no or very low $\lambda$)—outflows do not occur. The Bondi-type or smooth (super-hot) flow is generated and illustrated in Fig. \ref{Fig3}, which produces no outflows, and the simulation box is filled with inflows, representing a quasi-spherical accretion flow due to weaker rotational forces. The angular momentum distribution mostly depends on the temperature of inflowing gases at the outer boundary (see details in Fig. \ref{Fig6}). The smooth-cold (\ref{Fig4}a) has the highest $\lambda$ distribution in the advective flows, as shown in Fig. \ref{Fig6}, and it is generated with lower thermal energy than the other advective flows. It again has no or weak outflows. However, the smooth-hot (\ref{Fig4}b) has stronger outflows than the smooth-cold. The smooth-hot has $\lambda$ distribution in the middle range (between smooth-cold and smooth super-hot or Bondi-type flow). The smooth-hot is generated with higher thermal energies than the smooth-cold, and the smooth-hot has the stronger outflows due to the combined effect of the centrifugal and thermal forces. So, one can conclude that the nature of accretion flows is dependent on the local competitive forces in the flows. These advective flows are generated from the different regions of the OBC-plane. So we can conclude that the outflows in the disc can be dependent on the OBCs. The Keplerian disc, or non-advective accretion flow, falls outside the scope of this study due to model assumptions and, therefore, cannot be produced here.

Cases V and VI are plotted in panels (\ref{Fig4}c) and (\ref{Fig4}d), respectively, and both accretion flows exhibit shocks. Panel (\ref{Fig4}c) is derived from the OBC in the cold-mode region, referred to as shocked-cold, while the accretion solution of a panel (\ref{Fig4}d) is termed shocked-hot. The shocked-hot solution features stronger outflows in the post-shock region compared to the shocked-cold solution, which is visible in the figures through the velocity vectors in the post-shock region. These outflows originate near the black hole and do not span the entire disc, resembling the size of the observed jet base \citep{jbl99}. Thus, these outflows may serve as a precursor to jet formation if they are collimated and accelerated in the $z$-direction by an appropriate mechanism, such as radiative or magnetic acceleration and collimation \citep{c05,kcm14,vkm15,gbc20,kct24}. The post-shock region of the shocked-cold solution (panel \ref{Fig4}c) shows very weak outflows, which do not exit the simulation box, likely due to a weak shock and insufficient thermal pressure along the $z$-direction. The decreasing disc height in the pre-shock region of the shocked-cold flow indicates weak thermal pressure or stronger centrifugal force. In contrast, the shocked-hot flow maintains an almost constant disc height in the pre-shock region. Unlike the steady shock in the shocked (super-hot) case (Fig. \ref{Fig8R}), both cases have the quasi-steady shock location and accretion flow.  
Overall, we find that the solutions related to the hot mode OBCs can exhibit strong outflows or jet-like features in the post-shock region, unlike those with cold mode gas inflow. Thus, we can again conclude that OBCs are a crucial component in the study of discs around BH XBs, AGNs, and TDEs. Furthermore, { this study can allow us to predict about TDEs with jet-like and non-jet-like features in the accretion flow due to the effect of OBCs, assuming the TDEs have a hot flow component}. 
The three results of Fig. \ref{Fig4} are also regenerated with higher resolutions to check the consistency of the simulation results, and details are presented in the Appendix \ref{sec:hr}. 

Figure \ref{Fig6} shows the specific angular momentum ($\lambda$) distribution corresponding to five types of solutions, which encompass the $\lambda$ of the four solutions illustrated in Figure \ref{Fig4} and smooth (super-hot) flow or Bondi-type flow in Fig. \ref{Fig3}. All five cases share the same $\rob=400\rg$ but differ in \( \thob \) or $\lmdob$. Each case is plotted with a different line colour and mentioned with the name in Fig. \ref{Fig6}. The smooth-cold, smooth-hot, and smooth (super-hot) solutions are generated with the same Keplerian ($\lmdob=\lmk$) inflowing gas at $\rob$ but different $\thob$. The shock-cold and shock-hot solutions are generated with the sub-Keplerian ($\lmdob<\lmk$) inflowing gases at $\rob$. However, all these solutions are plotted up to the simulation boundary $\rinj=200\rg$.
We found that the smooth-cold solution exhibits the highest $\lambda$ distribution, while the smooth (super-hot) flow presents the lowest. This can be understood through the form of the viscous stress tensor ($\tau$) which is proportional to the temperature ($\Theta$) of the inflowing gases, implying \( \tau \propto \Theta \); moreover, angular momentum transport is most effective in the outer region (sub-sonic area) of the accretion flow \citep{ky21,k24}. Since the smooth-cold flow has the lowest $\thob\sim1.0e-3\thvir$, it consequently has the highest $\lambda$ distribution. Conversely, the smooth (super-hot) flow has the highest $\thob\gsim\thvir$, and shows more effective angular momentum transport yet has the lowest $\lambda$ distribution. Other solutions display $\lambda$ distributions lying between those of smooth-cold and smooth (super-hot) flows. This tendency has also been observed in semi-analytical studies \citep{k24,ky21}. Furthermore, Figure \ref{Fig6} illustrates that the high $\lambda$ flows (such as smooth-cold or ADAF-types) and low $\lambda$ flows (such as smooth (super-hot) or Bondi-type) are incapable of producing shocks in simulations, thus aligning with the conclusions from semi-analytical studies \citep{nkh97,lgy99,kc13,kc14,ck16,ky21,r25}. 

\section{SUMMARY AND DISCUSSION}
\label{sec:sumup}
\begin{table}
\centering
\caption{Types of the solutions with their OBCs domain in the OBC-plane and detected some prominent features.
Here, $c_{s_{vir}}$ is sound speed corresponding to $\thvir=k_B\tvir/(m_ec^2)$ at $\rob$.
}
\begin{tabular}{c c c c c c}
\hline \hline
Case No.&Type of solutions&OBCs&Outflows&Ang.Mom. distribution\\
\\
\hline
I&Smooth (super-hot) & $\be>0, \lmdob=\lmk, c_s|_{ob}>c_{s_{vir}}$&No outflows& Lowest among flows\\
\hline
II&Shocked (super-hot) & $\be>0, \lmdob=\lmk, c_s|_{ob}>c_{s_{vir}}$&Strong jet-like&In between\\
\hline
III&Smooth-cold & $\be<0, \lmdob=\lmk, c_s|_{ob}<< c_{s_{vir}}$&No or weak winds&Highest among flows\\
\hline
IV&Smooth-hot & $\be>0, \lmdob=\lmk, c_s|_{ob}<c_{s_{vir}}$&Strong winds &In between\\
\hline
V&Shocked-cold & $\be<0, \lmdob<\lmk, c_s|_{ob}<< c_{s_{vir}}$&Weak/failed jet-like &In between\\
\hline
VI&Shocked-hot  & $\be>0, \lmdob<\lmk, c_s|_{ob}<c_{s_{vir}}$&Strong jet-like &In between\\
\hline \hline
\end{tabular}
\label{tab:tab1}
\end{table}
In this simulation study, we investigated accretion flow using various OBCs and connected them to an OBC-plane (Fig. \ref{Fig1}). The OBC-plane can represent the physical OBCs of the sources. 
The OBC-plane is divided into cold- and hot-mode regions based on the behaviour of variation of the $\rob$ with $\be$. In the cold-mode region, the local energy of the inflowing gas has \( \be < 0 \) for any \( \lmdob \), whereas in the hot-mode region, it has \( \be > 0 \) for any \( \lmdob \). The hot-mode region is further divided into sub-virial and super-virial regions. So, the OBC-plane is divided into three regions: the cold-mode region ($\be<0$), sub-virial hot-mode region ($\be>0$), and super-virial hot-mode region ($\be>0$). Each region of the OBC-plane can generate smooth and shocked accretion solutions.
So, we selected six cases from the three areas of the OBC-plane based on the semi-analytical studies, and the simulation results were generated corresponding to those OBC points with the help of the semi-analytical solutions. We have summarised six cases including their OBCs and some typical natures in a table \ref{tab:tab1}. 
In the cold-mode region, we examined two types of solutions based on the initial Keplerian and sub-Keplerian angular momentum at the outer boundary, referred to as cases III and V in Table \ref{tab:tab1}. 
The remaining four cases are taken from the hot-mode region, based on the nature of solutions with varying initial temperature and initial angular momentum at the outer boundary.
Interestingly, the simulation results closely match the semi-analytical solutions associated with their respective OBCs. We found that angular momentum transportation is more effective in the outer part of the disc (or sub-sonic region), and it increases with increasing temperature at the outer boundary. 
Based on the obtained solutions with two types of inflowing gases (\cm and \hm) and their properties, we have made notable observations that could enhance our understanding of the accretion flow and also expect some relation of them with some observed characteristics of accreting astrophysical objects. 
The results of present study are as follows:

1. For the generation of the simulation results, we have used the initial input parameters at the simulation boundary from the semi-analytical solutions. Interestingly, the simulation results have nicely followed the semi-analytical solutions without simulating solutions with a full size of the theoretical outer boundary, see Figs (
\ref{Fig3}, \ref{Fig8R}). Moreover, in the present study, we found that the qualitative nature of the accretion flows is independent of outer boundary location, as we found all types of accretion flows in Figs. \ref{Fig4} and \ref{FigA} with different outer boundary locations, which again follows a result of the semi-analytical studies \citep{ky21,k24}. Although the quantitative things can change with a change in the outer boundary location. From this study, we can conclude that a large (actual size of disc) simulation box may not be necessary for studying the effect on the inner disc structure due to changes in the OBCs! 

2. We found that the accretion flows can be either shocked or smooth, depending on the OBCs.  
Since there is no control over the OBCs around the BHs. Therefore, we believe that models predicting only shocked or smooth accretion flows can not capture the complete physics of the accretion discs, such as the sandwich geometry model based on shocked advective flow \citep{ct95,c16} and the two-zone radial geometry model \citep{emn97} based solely on smooth advective flow. So, we believe that a disc model should be based on both the qualitative and quantitative nature of the inflowing gases. Moreover, we discovered that the steady and non-steady nature of the accretion flows can also depend on the OBCs. The disc’s non-steady characteristics may manifest as variabilities in the observed light curve, such as a quasi-steady shock in the disc.

3. We found that the advective solutions corresponding to the \hm gas can produce stronger outflows compared to those corresponding to the \cm gas in the OBC-plane. We investigated six cases from different regions of the OBC-plane and typically identified two types of accretion flows—smooth and shocked—within each region. Interestingly, the nature of all the simulation results is closely matched as predicted in the semi-analytical studies, such as the shocked (super-hot) flow in the simulation has almost the same location of the shock as in the semi-analytical solution (see Fig. \ref{Fig8R}), the smooth-cold flow has no shock as expected in the semi-analytical studies and other cases. The smooth solutions in this simulation study are named smooth-cold, smooth-hot, and smooth (super-hot) based on their OBC locations in the OBC-plane, which are similar to the ADAF, ADAF-thick, and Bondi types of flows in the semi-analytical studies \citep{nkh97,lgy99,ky21,k24}, respectively. 

4. The nature of the {general advective} flows predominantly relies on the {angular momentum} distribution, which is primarily influenced by the OBCs. The smooth flows can exhibit the highest {angular momentum} distribution (like smooth-cold solution) or the lowest {angular momentum} distribution (like smooth super-hot solution). In contrast, the shocked accretion flows maintain an angular momentum distribution that falls between the highest and lowest angular momentum of the smooth flows. For instance, if \( \lmdob = \lmk \) for the inflowing gases at some $\rob$ and where \( \thob \) is a free parameter that varies from very low temperature (\( << \thvir \)) to very high temperature, while keeping other model parameters fixed, we can derive all types of accretion flows listed in the table merely by increasing \( \thob \). Hence, the initial temperature of the inflowing gases is also a crucial parameter in the study of accretion flows. Furthermore, the initial angular momentum of the inflowing gas at the outer boundary is another significant parameter, as discussed in subsection \ref{subsec:vobcs}.  For example, if the OBCs are close to the dotted red line in panel (b) of Fig. \ref{Fig1}, then we get a smooth accretion flow, and if the OBCs are close to the solid-violet line, then the generated disc can have a shock.

5. Some observed accreting BHs exhibit jet on/off situations and sources with jets or no jets. 
For instance, { this study can allow us to predict about} the accretion disc of the TDEs with outflows/jets may be primarily influenced by the \hm inflowing gases compared to the disc of the TDEs with no outflows/jets, { if it has hot accretion flow component}. However, we believe that these scenarios can be better understood through a detailed study of the OBCs { with radiative dissipations}.

6. Some general advective flows demonstrate winds/outflows from the disc, as shown in Figs. \ref{Fig4} and \ref{Fig8R}. These have outflow velocities exceeding $0.05c$ in the box and increase along the $z$-axis. These outflow velocities have been measured in some AGNs \citep{ppf17}. However, these velocities are lower than the observed ultra-fast outflows in AGNs. Nonetheless, we believe that these velocities could be higher due to the increasing trend of the jet velocities along the $z$-direction. Moreover, some accretion flows exhibit failed outflows, as illustrated in Fig. \ref{Fig4}(c). Moreover, the outflow velocities can be increased with the inclusion of the magnetic field in the flow, as seen in the MHD study \citep{gbc20}. 

{ These conclusions based on the assumption of radiative inefficient hot accretion flows and they may very if the flow is super-Eddington or radiative efficient or both. This study can be considered as pilot project study because we need a comprehensive model which can explain the most of the properties of the accreting sources, not piece by piece study of the properties of a source.}
As seen in observational studies, the shape of the HID diagram in the BXBs deviates from the typical $q$-shape, and this shape can vary from outburst to outburst. Additionally, the luminosities and jet strengths (strong/weak jets) can also fluctuate between outbursts in the same sources. Therefore, we believe that these variations may stem from the qualitative and quantitative nature of the inflowing gases at the outer boundary, as the nature of the inflowing gases can influence the domination of disc components in disc models and characteristic timescales of the outbursts of the sources, as predicted in this study \citep{ky21}. 
This study has concentrated on the qualitative aspects of the inflowing gases, presenting them on the OBC-plane. 
Since the simulation results closely match the semi-analytical results. Therefore, we can conclude that a 2D disc model based on both qualitative and quantitative inflow gases constructed through the semi-analytical numerical methods as described in \cite{kg18}, can accurately represent the steady properties of accreting systems. This 2D semi-analytical model can be developed for any scale, making it less time-consuming and resource-intensive than the alternative simulations.
However, this type of study needs many improvements in future studies, some of which are briefly discussed in the following paragraphs.

{\it Improvements Required:}
This simulation study primarily focuses on investigating various types of accretion solutions from different regions of the OBC plane. However, it lacks detail regarding certain properties of accretion flows with varying viscosity parameters and the incorporation of appropriate radiative processes and magnetic fields. Consequently, in the next study, we aim to reproduce some of the results in detail and conduct a comparative analysis to explore further physics and applications, including more physical processes. This includes examining reasons for steady and quasi-steady accretion flows, as well as steady/non-steady shocks resulting from changes in the OBCs,  details of generation and no generation of the outflows (under preparation), and analysing non-steady shocks for QPO study. We will also revisit the properties of the accretion flows by integrating suitable radiative processes in future work, which will provide a better understanding of the accretion flow both qualitatively and quantitatively as OBCs change.

Furthermore, this study is conducted with a single type of inflowing gas at the outer boundary to produce one type of solution. However, future research could include simulations with a mixture of gases at the boundary, such as combining hot and cold gases or mixing Keplerian and sub-Keplerian gases, among other combinations. These scenarios could better reflect more realistic boundary conditions of the inflowing gases, leading to a deeper understanding of the accretion flow physics.

From the above study, we can conclude that accretion physics encompasses the dynamics of competing forces, leading to both smooth and shocked accretion flows with or without generating outflows. We found that the thermal and centrifugal forces are the primary competing forces against central gravity. Furthermore, both forces help generate outflows if both forces have close comparison based on the found disc structures in this study; for instance, smooth-cold flows possess the highest angular momentum distribution among general advective flows, yet both thermal and high centrifugal force are insufficient to produce outflows. Conversely, smooth (super-hot) flows have higher thermal energy but the lowest angular momentum distribution among the advective flows, indicating that the combined forces are inadequate for generating outflows within the disc. However, the advective solutions with the intermediate values of thermal and angular momentum distribution have outflows, such as the smooth-hot and shocked flows. Notably, both the thermal and angular momentum distributions in the accretion flows are highly dependent on the OBCs or {physical properties of the external gases}.
Replicating actual physical OBCs based on the type of sources in the semi-analytical and simulation studies is essential yet challenging. Nevertheless, we have attempted to address these aspects in this simulation study. We predicted that the nature of the disc structure can depend on the type of dominant OBCs of the inflowing gases in the accreting sources.
We believe that astrophysics are about the possibilities, at least with the present status of knowledge about the universe. The possibilities can come from the study of the theoretical and numerical models. 

\section*{Acknowledgements}
This work was supported by the National Research Foundation of Korea (NRF) grant funded by the Korea government (MSIT) (No. RS-2023-00277151). 
We thank the anonymous referee for comments and suggestions that have improved this work.

\section*{Data Availability}
The data underlying this article will be available upon reasonable request. 



\bibliographystyle{mnras}
\bibliography{THAF} 

\appendix
%
%
%
\section{Calculation of the initial input conditions for simulation}\label{sec:tsss}
We have used local flow variables of the semi-analytical accretion solutions as the initial conditions for the generation of the simulation results. To obtain the semi-analytical solutions, we assumed that the disc is axisymmetric ($\partial/\partial\phi=0$), steady-state ($\partial/\partial t=0$), and in vertical equilibrium ($\vz=0$) along the $ z$-axis in the cylindrical geometry.  
After imposing these assumptions on Eqs. (\ref{mc.eq}-\ref{ec.eq}), Only the radial derivatives of the flow variables are left. 
Now, the equations of motion for steady-state flow on the equatorial plane with the following \cite{k24,kc13} can be written as,
the integrated form of the continuity equation (\ref{mc.eq}) gives the mass-accretion rate ($\dot{M}$) equation,
\begin{equation}
\frac{1}{r}\frac{d(r\rho\vr)}{dr}=0, ~\Rightarrow ~\dot{M}=4\pi\rho Hr\ve, 
\label{mdot.eq}
\end{equation}
The z-component of the momentum equation (\ref{mcz.eq}), assuming vertical hydrostatic equilibrium in the z-direction of the disc, gives the local disc half-height expression,
\begin{equation}
H=\sqrt{\frac{2r}{\gamma}}c_s(r-1)
\label{hh.eq}
\end{equation}
The other components of the momentum conservation equations (Eqs. \ref{mcr.eq} and \ref{mcp.eq}) and the energy generation equation still have radial derivatives in their equations. We now simplify these equations together with the help of Eqs. (\ref{mdot.eq} and \ref{hh.eq}) and separated out the radial derivatives of the flow variables, which can be written as,
\begin{eqnarray}
\frac{d\ve}{dr}=\frac{\frac{\ve}{\gamma+1}\frac{(5r-3)}{r(r-1)}-\frac{(\lambda_K^2-\lambda^2)\ve}{c_s^2r^3}+\gamma^2\left(\frac{\gamma-1}{\gamma+1}\right)\frac{\ve^2\lambda_K(\lambda-\lambda_0)^2}{\alpha c_s^4r^4}}{\frac{\ve^2}{c_s^2}-\frac{2}{\gamma+1}}, \label{ssdv.eq}\\
\frac{d\lambda}{dr}=\frac{2\lambda}{r}-\frac{\gamma\ve\lambda_K(\lambda-\lambda_0)}{\alpha c_s^2r^2}, \label{ssdl.eq} ~\mbox{and}~~~~~~~~~~~\\
\frac{dc_s}{dr}=\left(\frac{c_s}{\ve}-\frac{\gamma\ve}{c_s}\right)\frac{d\ve}{dr}+\frac{(5r-3)c_s}{2r(r-1)}+\frac{\gamma(\lambda^2-\lambda_K^2)}{c_sr^3},
\label{ssdcs.eq}
\end{eqnarray}
here, inflow velocity ($\ve$), sound speed ($c_s$) and specific angular momentum ($\lambda$) are defined on the equatorial plane. We solved these three differential equations using the 4th-order Runge-Kutta method; details of the numerical methodology are provided in \cite{kc13}. 
We have surveyed the entire OBC plane and widely identified two main types of solutions: shocked and smooth accretion flows. 
All these solutions have already been mentioned in different semi-analytical studies with/without fully exploring OBCs \citep{c89,nkh97,lgy99,y99,ky21,k24}. The smooth solutions can be further categorised based on their disc thickness and the location of critical points, including ADAF \citep{nkh97} and ADAF-thick \citep{lgy99}, both of which pass through an inner critical point (close to BH), while Bondi-type flow \citep{B52,c89} passes through an outer critical point (far from the BH). Interestingly, our {semi-analytical} studies indicate that all these solution types are qualitatively independent of $\rob$ \citep{ky21,kg19}. 
All these solutions originate from the different regions of the OBC-plane. The Bondi-types and shock solutions are obtained from the super-virial region of the OBC-plane, and others are from the sub-virial regions. 
We used flow variables from these solutions at a radius as the initial input conditions in the simulation for generating the accretion flows in this study. 
%
\section{Accretion flows with higher resolution}
\label{sec:hr}
\begin{figure*}
\centering
\begin{subfigure}{.45\textwidth}
		\includegraphics[width=\textwidth]{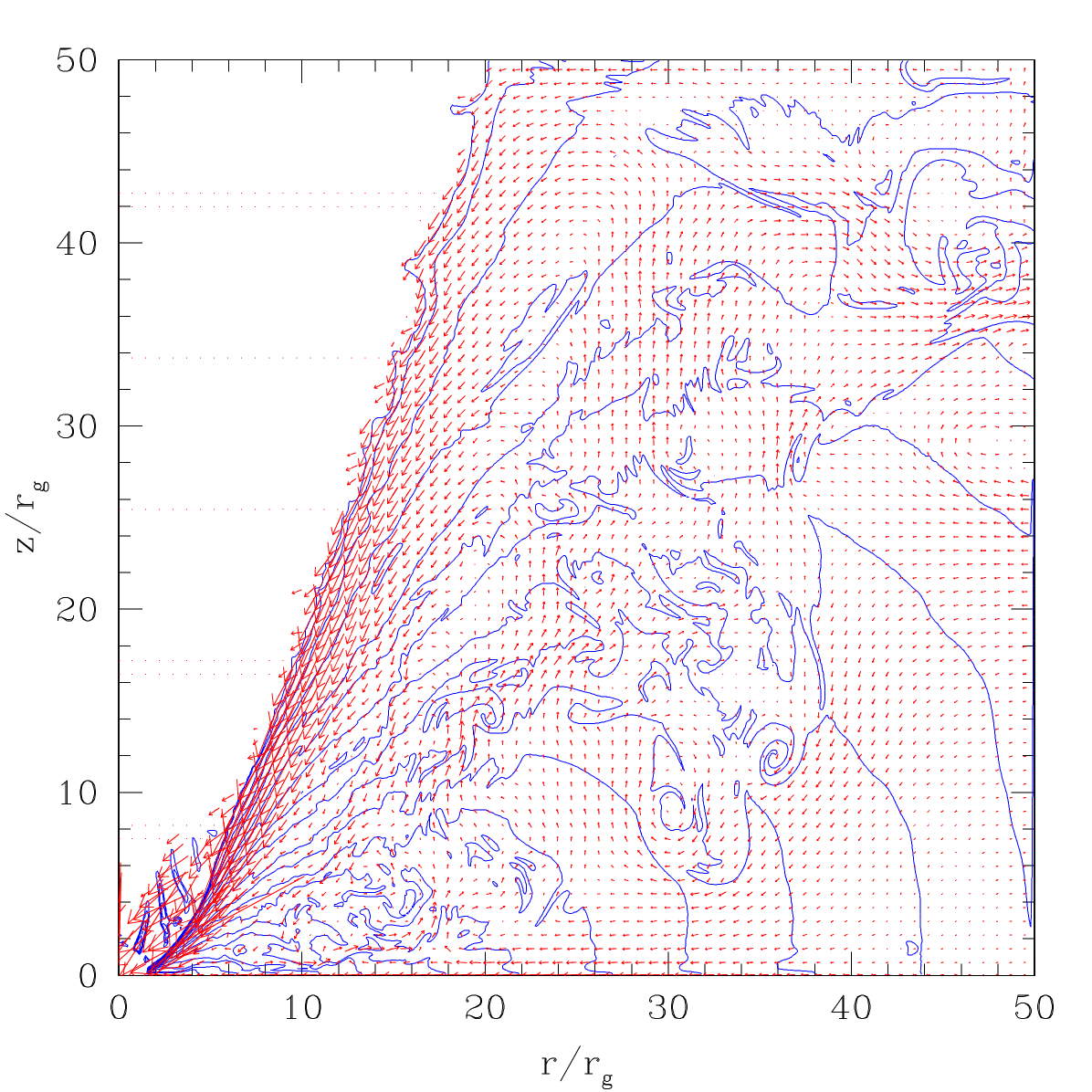}
		\caption{Case III - Smooth-cold}
	\end{subfigure}
	\begin{subfigure}{.45\textwidth}
		\includegraphics[width=\textwidth]{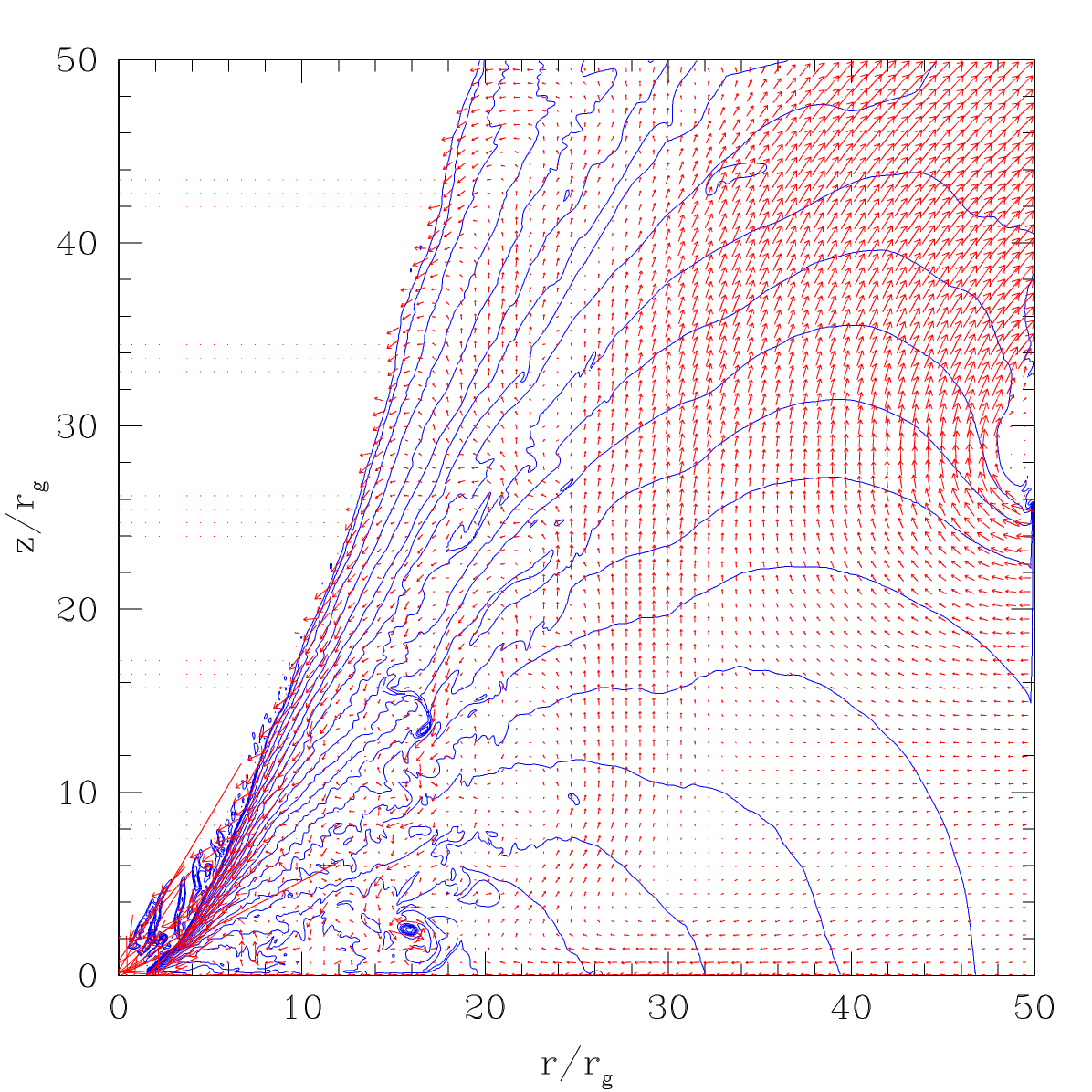}
		\caption{Case IV - Smooth-hot}
	\end{subfigure}
	\begin{subfigure}{.47\textwidth}
		\includegraphics[width=\textwidth]{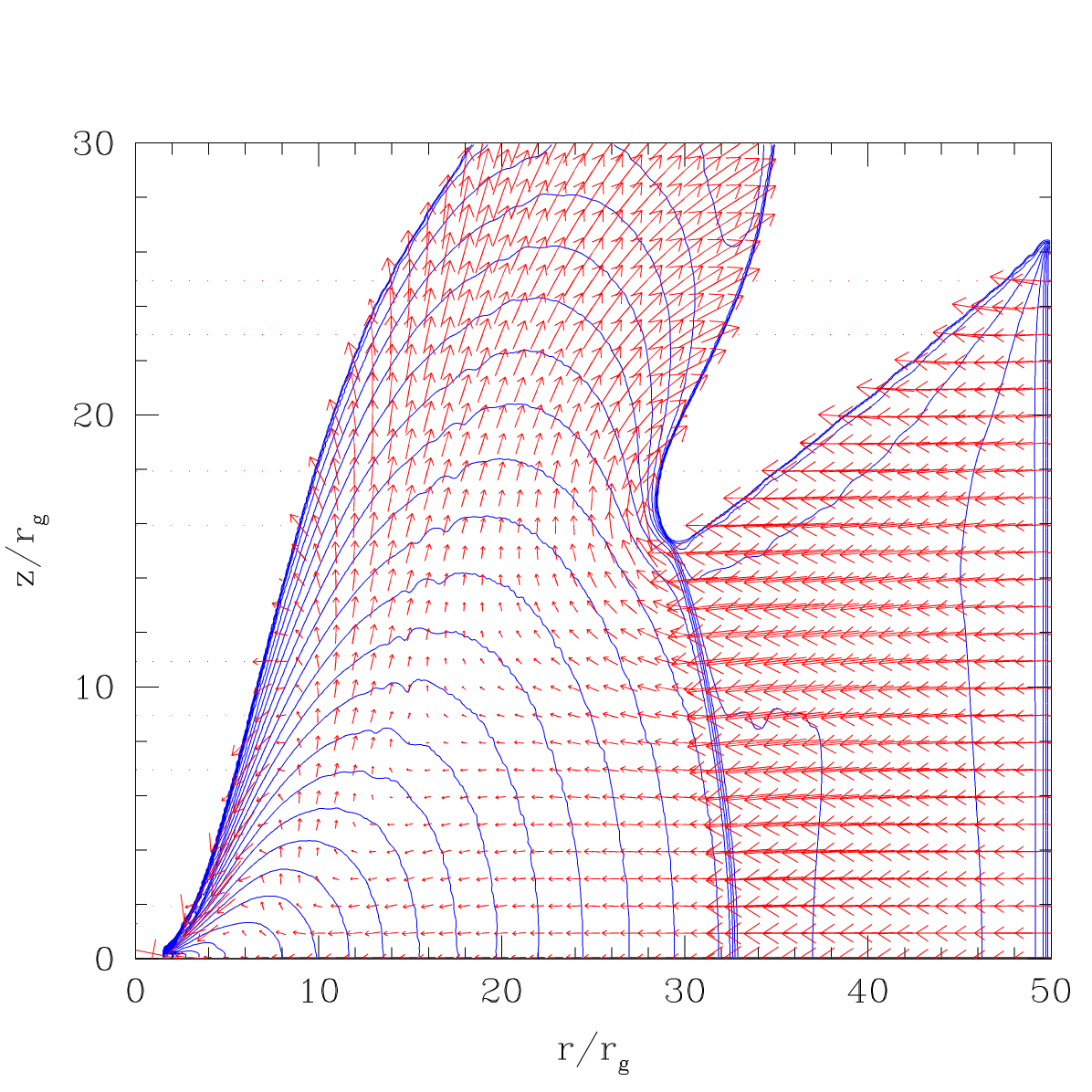}
		\caption{Case V - Shocked-cold}
	\end{subfigure}
	\begin{subfigure}{.43\textwidth}
		\includegraphics[width=\textwidth]{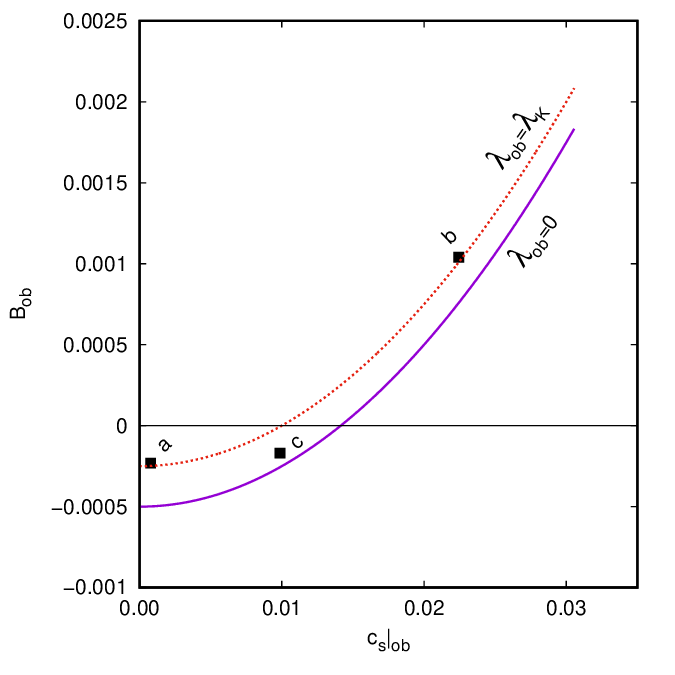}
		\caption{}
	\end{subfigure}
  \caption{Panel (a), Panel (b), and Panel (c) represent smooth-cold,  smooth-hot, and shocked-cold solutions, respectively. The simulation BCs at $\rinj=50\rg$ are $\vinj=-8.662552\times10^{-4}c, \csinj=6.154559\times10^{-2}c$, and $\lmdinj=3.488979\rg c$ 
for panel (a), $\vinj=-1.215449\times10^{-3}c, \csinj=6.469556\times10^{-2}c$, and $\lmdinj=3.062185\rg c$ 
for panel (b), and $\vinj=-5.424110\times10^{-3}c, \csinj=6.254159\times10^{-2}c$, and $\lmdinj=1.971723\rg c$ 
for panel (c). The simulation BCs at $\rinj$ are obtained from the semi-analytical solutions corresponding to the OBCs at $\rob=2400\rg$, which are presented in panel (d). All three cases are generated from the same types of OBCs in the OBC-plane as generated in the panels (\ref{Fig4}a - \ref{Fig4}c) with corresponding OBCs at $\rob=400\rg$ in panel (b) of Fig.\ref{Fig1}. 
} 
\label{FigA}
\end{figure*}

To check the consistency of the simulation results with resolution, we have regenerated disc structures of the Fig. \ref{Fig4} with higher resolutions, $600\times600$ cells in a $(50\rg\times50\rg)$ or $(50\rg\times30\rg)$ box size.  Moreover, we have also tested a prediction of semi-analytical studies about the generation of types of accretion solutions that are independent of $\rob$, so we have chosen $\rob=2400\rg$ here. We have chosen the same types of OBCs at $\rob=2400\rg$ in panel (d) of Fig. \ref{FigA} as in panel (b) of Fig. \ref{Fig1}, which can generate the same types of accretion flows. 
In Fig. \ref{FigA}, we have plotted three simulation results: two from the cold-mode region, which have smooth (panel a) and shocked (panel c) flow, and the other from the hot-mode region of the OBC-plane, which has smooth (panel b) flow. 
Interestingly, we have found similar types of disc structures in panels of Fig. \ref{FigA} as in panels (a, b, and c) of Fig. \ref{Fig4}, which are named smooth-cold, smooth-hot, and shocked-cold flows.  The results of Figs. (\ref{Fig4} with $\rob=400\rg$ and \ref{FigA} with $\rob=2400\rg$) also convey a message about the nature of accretion flow being independent of the outer boundary locations, which follows one more prediction of the semi-analytical studies. Although the quantitative things (such as shock strength and location, outflow strength, and so on) of the accretion flows can vary with the change in the outer boundary location.

\bsp	
\label{lastpage}
\end{document}